\documentclass[11pt,draftcls,onecolumn]{IEEEtran}

\usepackage{amsmath,amsfonts,amssymb,amsthm}
\usepackage{mathrsfs}
\usepackage{comment,blkarray}
\usepackage{multirow,bigdelim}
\usepackage{cite}
\usepackage[ruled,commentsnumbered, vlined]{algorithm2e}
\usepackage{tikz}
\usetikzlibrary{arrows,automata}
\usepackage[latin1]{inputenc}
\usepackage{verbatim}
\usepackage{graphicx}
\usepackage{subfigure}		
\usepackage{booktabs}
\usepackage{caption}													
\usepackage{cases}

\makeatletter

\newcommand{\Rmnum}[1]{\expandafter\@slowromancap\romannumeral #1@}

\newif\if@borderstar
\def\bordermatrix{\@ifnextchar*{%
  \@borderstartrue\@bordermatrix@i}{\@borderstarfalse\@bordermatrix@i*}%
}
\def\@bordermatrix@i*{\@ifnextchar[{\@bordermatrix@ii}{\@bordermatrix@ii[()]}}
\def\@bordermatrix@ii[#1]#2{%
\begingroup
  \m@th\@tempdima8.75\p@\setbox\z@\vbox{%
    \def\cr{\crcr\noalign{\kern 2\p@\global\let\cr\endline }}%
    \ialign {$##$\hfil\kern 2\p@\kern\@tempdima & \thinspace %
    \hfil $##$\hfil && \quad\hfil $##$\hfil\crcr\omit\strut %
    \hfil\crcr\noalign{\kern -\baselineskip}#2\crcr\omit %
    \strut\cr}}%
  \setbox\tw@\vbox{\unvcopy\z@\global\setbox\@ne\lastbox}%
  \setbox\tw@\hbox{\unhbox\@ne\unskip\global\setbox\@ne\lastbox}%
  \setbox\tw@\hbox{%
    $\kern\wd\@ne\kern -\@tempdima\left\@firstoftwo#1%
    \if@borderstar\kern2pt\else\kern -\wd\@ne\fi%
    \global\setbox\@ne\vbox{\box\@ne\if@borderstar\else\kern 2\p@\fi}%
    \vcenter{\if@borderstar\else\kern -\ht\@ne\fi%
    \unvbox\z@\kern -\if@borderstar2\fi\baselineskip}%
    \if@borderstar\kern -2\@tempdima\kern2\p@\else\,\fi\right\@secondoftwo#1 $%
  }\null \;\vbox{\kern\ht\@ne\box\tw@}%
\endgroup
}
\makeatother

\newtheorem{thm}{Theorem}

\newtheorem{lemma}[thm]{Lemma}
\newtheorem{eg}{Example}
\newtheorem{prop}{Proposition}

\newtheorem{defn}{Definition}
\newtheorem{rem}[thm]{Remark}

\allowdisplaybreaks

\newcommand{\w}{{\omega}}
\newcommand{\vb}{\vec{b}}
\newcommand{\vc}{\vec{c}}

\newcommand{\vf}{\vec{f}}

\newcommand{\vv}{\vec{v}}
\newcommand{\vu}{\vec{u}}
\newcommand{\vh}{\vec{h}}

\newcommand{\Fq}{\mathbb{F}_q}
\newcommand{\mB}{\mathcal{B}}
\newcommand{\mA}{\mathscr{A}}
\newcommand{\mE}{\mathscr{E}}
\newcommand{\mC}{\mathcal{C}}
\newcommand{\mL}{\mathcal{L}}
\newcommand{\mO}{\mathcal{O}}

\newcommand{\bzero}{{\vec{0}}}

\newcommand{\Rank}{{\mathrm{Rank}}}
\newcommand{\tail}{{\mathrm{tail}}}
\newcommand{\head}{{\mathrm{head}}}
\newcommand{\Out}{{\mathrm{Out}}}
\newcommand{\In}{{\mathrm{In}}}

\newcommand{\CUT}{{\mathrm{CUT}}}

\begin{document}

\title{Local-Encoding-Preserving Secure Network Coding---Part \Rmnum{2}: Flexible Rate and Security~Level
}

\author{Xuan~Guang,~\IEEEmembership{Member,~IEEE,}
        ~Raymond~W.~Yeung,~\IEEEmembership{Fellow,~IEEE,}
        ~and~Fang-Wei~Fu,~\IEEEmembership{Member,~IEEE}
}

\markboth{}%
{}
%


\maketitle

\begin{abstract}
In this two-part paper, we consider the problem of secure network coding when the information rate and the security level can change over time. To efficiently solve this problem, we put forward local-encoding-preserving secure network coding, where a family of secure linear network codes (SLNCs) is called {\em local-encoding-preserving} if all the SLNCs in this family share a common local encoding kernel at each intermediate node in the network. In the current paper (i.e., Part~\Rmnum{2} of the two-part paper), we first consider the design of a family of local-encoding-preserving SLNCs for a fixed rate and a flexible security level. We present a novel and efficient approach for constructing upon an SLNC that exists a local-encoding-preserving SLNC with the same rate and the security level increased by one. Next, we consider the design of a family of local-encoding-preserving SLNCs for a fixed dimension (equal to the sum of rate and security level) and a flexible pair of rate and security level. We propose another novel approach for designing an SLNC such that the same SLNC can be applied for all the rate and security-level pairs with the fixed dimension. Also, two polynomial-time algorithms are developed for efficient implementations of the two proposed approaches, respectively. Furthermore, we prove that both approaches do not incur any penalty on the required field size for the existence of SLNCs in terms of the best known lower bound by Guang and Yeung. Finally, we consider the ultimate problem of designing a family of local-encoding-preserving SLNCs that can be applied to all possible pairs of rate and security level. By combining the construction of a family of local-encoding-preserving SLNCs for a fixed security level and a flexible rate (which has been obtained in Part~\Rmnum{1}~\cite{part1}) with the constructions of the two families of local-encoding-preserving SLNCs in the current paper in suitable ways, we can obtain a family of local-encoding-preserving SLNCs that can be applied for all possible pairs of rate and security level. Three possible such constructions are presented.
\end{abstract}

%
\IEEEpeerreviewmaketitle

\section{Introduction}

As discussed in Part~\Rmnum{1}~\cite{part1} of this two-part paper, in a secure network coding system, the requirements for information transmission and information security may vary, and so the information rate and the security level of the system may need to be chosen differently at different times. Thus we are motivated to put forward {\em local-encoding-preserving secure network coding}, where a family of secure linear network codes (SLNCs) is called {\em local-encoding-preserving} if all the SLNCs in this family share a common local encoding kernel at each intermediate node in the network. In other words, regardless of which SLNC in this family to be used, the same local encoding kernel at each intermediate node is applied for local encoding. Further, we considered the design of a family of local-encoding-preserving SLNCs for a fixed security level and a flexible rate. 
Specifically, we proposed an approach to constructing a family of local-encoding-preserving SLNCs with a fixed security level $r$ ($0\leq r \leq C_{\min}$) and multiple rates from $0$ to $C_{\min}-r$, the allowed maximum rate, where $C_{\min}$ is the smallest minimum cut capacity between the source node and the sink nodes. We also developed a polynomial-time algorithm for efficient implementation of our approach. Furthermore, it was proved that our approach does not incur any penalty on the required field size for the existence of SLNCs in terms of the best known lower bound in~\cite{GY-SNC-Reduction}.

In the current paper (i.e., Part~\Rmnum{2} of the two-part paper), we continue the studies in Part~\Rmnum{1} by considering the design of a family of local-encoding-preserving SLNCs for a fixed rate and a flexible security level. Our approach is totally different from all the previous approaches used in related problems (including the approach in Part~\Rmnum{1})~\cite{Fong-Yeung-variable-rate, Guang-uni-MDS,part1}. Upon an SLNC that exists, we construct a local-encoding-preserving SLNC with the same rate and at one security level higher, which leads to an increase of the code dimension (equal to the sum of rate and security level). In contrast, in all the previous approaches, the dimension of the newly constructed linear network code decreases.

Specifically, we start with an SLNC with the required rate, denoted by $\w$, and the security level $0$ (which in fact is an $\w$-dimensional linear network code), and apply the proposed approach repeatedly. As such, we can obtain a family of local-encoding-preserving SLNCs with the fixed rate $\w$ and multiple security levels from $0$ to $C_{\min}-\w$, the allowed maximum security level. Based on this approach, we further develop a polynomial-time algorithm for efficient implementation. Also, we prove that our approach incurs no penalty (even better for some cases) on the required field size for the existence of SLNCs in terms of the best known lower bound~\cite{GY-SNC-Reduction}.

Next, we consider the design of a family of local-encoding-preserving SLNCs for a fixed dimension and a flexible pair of rate and security level, i.e., all rate and security-level pairs such that the sum of the rate and the security level is equal to a constant. Toward this end, we use another novel approach to design an SLNC such that with the same SLNC, all the rate and security-level pairs with the fixed dimension are applicable. With this approach, we develop a polynomial-time algorithm for the construction of such an SLNC. Again, for our approach, there is no penalty on the field size of the existence of SLNCs in terms of the best known lower bound~\cite{GY-SNC-Reduction}.

Finally, we consider the design of a family of local-encoding-preserving SLNCs which can be applied to all possible pairs of rate and security level, i.e., all the nonnegative integer pairs $(\w,r)$ of rate $\w$ and security level $r$  with $\w+r\leq C_{\min}$. The set of all such the pairs forms the {\em rate and security-level region}. By combining the constructions of the 3 families of local-encoding-preserving SLNCs in Part~\Rmnum{1} and Part~\Rmnum{2} in suitable ways, we can construct a family of local-encoding-preserving SLNCs achieving all the pairs in the rate and security-level region. Three possible such constructions are presented.

The organization of this paper is as follows.
In Section~\ref{Sec_Preli}, we present secure network coding and the preliminaries, and introduce the necessary definitions and a proposition. In Sections~\ref{Sec_v-s-l} and~\ref{sec_fix_dim}, we design a family of local-encoding-preserving SLNCs for a fixed rate and a flexible security level, and a family of local-encoding-preserving SLNCs for a fixed dimension and a flexible rate and security-level pair. Section~\ref{Sec_ultimate} is devoted to the design of a family of local-encoding-preserving SLNCs achieving all the pairs in the rate and security-level region. We conclude in Section~\ref{Sec_conclusion} with a summary of our results.

\section{Preliminaries}\label{Sec_Preli}

For the completeness of this paper, in this section we briefly present secure network coding and introduce the necessary notation and definitions. We refer the reader to Section~\Rmnum{2} of Part~\Rmnum{1}~\cite{part1} of the current paper for more details.

We consider a finite directed acyclic network $G=(V, E)$ with a single source $s$ and a set of sink nodes $T\subseteq V\setminus \{s\}$, where $V$ is the set of nodes and $E$ is the set of edges of $G$. For a directed edge~$e$ from node $u$ to node $v$, the node $u$ is called the {\em tail} of $e$ and the node $v$ is called the {\em head} of $e$, denoted by $\tail(e)$ and $\head(e)$, respectively. Further, for a node $u$, let $\In(u)=\{e \in E:\ \head(e)=u\}$ and $\Out(u)=\{e\in E:\ \tail(e)=u\}$. Without loss of generality, assume that there are no incoming edges for the source node $s$ and no outgoing edges for any sink node $t\in T$. For convenience sake, however, we let $\In(s)$ be a set of $n$ {\em imaginary incoming edges}, denoted by $d_i$, $1\leq i \leq n$, terminating at the source node $s$ but without tail nodes, where the nonnegative integer $n$ is equal to the dimension of the network code in discussion.
This will become clear later (see Definition~\ref{def-LNC}). Then, we see that $\In(s)=\big\{d_i:~1\leq i \leq n \big\}$. An index taken from an alphabet can be transmitted on each edge $e$ in $E$. In other words, the capacity of each edge is taken to be $1$. Parallel edges between two adjacent nodes are allowed.

In a network $G$, a {\em cut} between the source node $s$ and a non-source node $t$ is defined as a set of edges whose removal disconnects $s$ from $t$. The \textit{capacity} of a cut between $s$ and $t$ is defined as the number of edges in the cut, and the minimum of the capacities of all the cuts between $s$ and $t$ is called the \textit{minimum cut capacity} between them, denoted by $C_t$. A cut between $s$ and $t$ is called a \textit{minimum cut} if its capacity achieves the minimum cut capacity between them. These concepts can be extended from a non-source node $t$ to an edge subset $A$ of $E$ (cf.~\cite[Section~\Rmnum{2}-A]{part1}).

It has been shown in \cite{Li-Yeung-Cai-2003, Koetter-Medard-algebraic} that {\em linear network coding} over a finite field is sufficient for achieving $C_{\min}\triangleq \min_{t\in T}C_t$, the theoretical maximum information rate for multicast~\cite{Ahlswede-Cai-Li-Yeung-2000}. The formal definition of a linear network code is given as follows.

\begin{defn}\label{def-LNC}
Let $\Fq$ be a finite field of order $q$, where $q$ is a prime power, and $n$ be a nonnegative integer. An $n$-dimensional $\Fq$-valued linear network code $\mC_n$ on the network $G=(V,E)$ consists of an $\Fq$-valued $\lvert \In(v)\rvert \times \lvert \Out(v)\rvert$ matrix $K_v=[k_{d,e}]_{d\in \In(v), e\in \Out(v)}$ for each non-sink node $v$ in $V$, i.e.,
\begin{align*}
\mC_n=\big\{ K_v:~v\in V\setminus T \big\},
\end{align*}
where $K_v$ is called the local encoding kernel of $\mC_n$ at $v$, and $k_{d,e}\in \Fq$ is called the local encoding coefficient for the adjacent edge pair $(d,e)$.
\end{defn}

For a linear network code $\mC_n$, the local encoding kernels $K_v$ at all the non-sink nodes $v\in V\setminus T$ induce an $n$-dimensional column vector $\vf^{\,(n)}_e$ for each edge $e$ in $E$, called the \textit{global encoding kernel} of~$e$, which can be calculated recursively according to a given ancestral order of edges in $E$ by
        \begin{align}\label{equ_ext_f}
        \vf^{\,(n)}_e=\sum_{d\in \In(\tail(e))}k_{d,e}\cdot\vf^{\,(n)}_d,
        \end{align}
with the boundary condition that $\vf_{d}^{\,(n)}$, $d\in \In(s)$ form the standard basis of the vector space $\Fq^n$. The set of global encoding kernels for all $e\in E$, i.e., $\big\{ \vf^{\,(n)}_e:~e\in E \big\}$, is also used to represent this linear network code $\mC_n$. However, we remark that a set of global encoding kernels $\big\{ \vf^{\,(n)}_e:~e\in E \big\}$ may correspond to more than one set of local encoding kernels $\big\{ K_v:~v\in V\setminus T \big\}$.

In using of this linear network code $\mC_n$, let $\mathbf{x}=\big(x_1 \ \ x_2 \ \ \cdots \ \ x_n \big) \in \Fq^{n}$ be the input of the source node~$s$. We assume that the input $\mathbf{x}$ is transmitted to $s$ through the $n$ imaginary incoming channels of the source node~$s$. Without loss of generality, $x_i$ is transmitted on the $i$th imaginary channel $d_i$, $1\leq i \leq n$. For each edge $e\in E$, we use $y_e$ to denote the message transmitted on $e$. Then $y_e$ can be calculated recursively by the equation
\begin{align}\label{equ_U_e}
y_e=\sum_{d\in \In(\tail(e))}k_{d,e}\cdot y_d
\end{align}
according to the given ancestral order of edges in $E$, with $y_{d_i}\triangleq x_{i}$, $1\leq i \leq n$. We see that $y_e$ in fact is a linear combination of the $n$ symbols $x_i$, $1\leq i\leq n$ of $\mathbf{x}$. It is readily seen that $y_{d_i}= \mathbf{x} \cdot \vf_{d_i}^{\,(n)}$\, (\,$=x_{i}$), $1\leq i \leq n$. Then it can be shown by induction via \eqref{equ_ext_f} and \eqref{equ_U_e} that
\begin{align}
y_e= \mathbf{x} \cdot \vf_{e}^{\,(n)}, \quad \forall\,e\in E.
\end{align}
Furthermore, for each sink node $t\in T$, we define the matrix $F_t^{(n)}=\left[\vf_e^{\,(n)}:~e\in \In(t)\right]$. The sink node~$t$ can decode the source message with zero error if and only if $F_t^{(n)}$ is full rank, i.e., $\Rank\big(F_t^{(n)}\big)=n$. We say that an $n$-dimensional linear network code $\mC_n$ is {\em decodable} if for each sink node $t$ in $T$, the rank of the matrix $F_t^{(n)}$ is equal to the dimension $n$ of the code. Next, we present the transformation of a linear network code.

\begin{prop}[\!\!{\cite[Theorem~1]{part1}}]\label{prop}
Let $\mC_n=\big\{ K_v:~v\in V\setminus T \big\}$ be an $n$-dimensional decodable linear network code over a finite field $\Fq$ on the network $G=(V,E)$, of which the global encoding kernels are $\vf^{\,(n)}_e$, $e\in E$. Let $Q$ be an $m\times n$ ($m\leq n$) matrix over $\Fq$ and $Q\cdot\mC_n = \big\{ K_v^{(Q)}:~v\in V\setminus T \big\}$ with $K_s^{(Q)}=Q \cdot K_s$ and $K_v^{(Q)}=K_v$ for all $v\in V\setminus (\{s\}\cup T)$.
Then $Q \cdot \mC_n$ is an $m$-dimensional linear network code over $\Fq$ on $G$, of which the global encoding kernels are $Q \cdot \vf^{\,(n)}_e$, $e\in E$. This linear network code $Q\cdot\mC_n$ is called the transformation of $\mC_n$ by the matrix $Q$. In particular, $Q\cdot\mC_n$ is decodable provided that $Q$ is full row rank, i.e., $\Rank\big(Q\big)=m$.
\end{prop}

Now, we present the secure network coding model. The source node $s$ generates a random source message $M$ taking values in the message set $\Fq^{\w}$ according to the uniform distribution, where the nonnegative integer~$\w$ is called the {\em information rate}. The source message $M$ needs to be multicast to each sink node $t\in T$, while being protected from a wiretapper who can access one but not more than one arbitrary edge subset of size at most~$r$, where the nonnegative integer $r$ is called the {\em security level}. The network $G$ with a required security level $r$ is called an $r$-{\em wiretap network}. To combat the wiretapper in our wiretap network model, a random \textit{key} generated at the source node is used to randomize the source message. This key is a random variable $K$ taking values in a set of keys $\Fq^r$ according to the uniform distribution.

We now consider a {\em secure linear network code} (SLNC) on an $r$-wiretap network $G$. Let $n=\w+r$, the sum of the information rate $\w$ and the security level $r$. An $\Fq$-valued $n$-dimensional SLNC on the $r$-wiretap network $G$ is
an $\Fq$-valued $n$-dimensional linear network on $G$ such that the following are satisfied:
\begin{itemize}
  \item {\em decoding condition}: every sink node is able to decode the source message $M$ with zero error;
  \item {\em security condition}: $I(Y_A ; M)=0$,\footnote{Here, $I(Y_A ; M)$ denotes the mutual information between $Y_A$ and $M$.} $\forall~A\subseteq E$ with $\lvert A \rvert\leq r$, where we denote by $Y_e$ the random variable transmitted on the edge $e$ that is a linear function of the random source message $M$ and the random key $K$, and denote $(Y_e: e\in A)$ by $Y_A$ for a subset $A\subseteq E$.
  \end{itemize}
The nonnegative integers $\w$ and $r$ are also referred to as the {\em information rate} and {\em security level} of the SLNC, respectively. When $r=0$, the secure network coding model reduces to the original network coding model.

Next, we specify a construction of SLNCs which will be used subsequently. Before specifying this code construction, we need several graph-theoretic concepts. By applying these concepts, the required field size for the existence of SLNCs can be reduced significantly. Let $A\subseteq E$ be an edge subset. A minimum cut between $s$ and $A$ is {\em primary} if it separates $s$ and all the minimum cuts between $s$ and $A$. Such a primary minimum cut is unique and can be found in polynomial time \cite{GY-SNC-Reduction}. Furthermore, we say an edge subset is {\em primary} if this edge subset itself is its primary minimum cut from $s$. We use $\mA_r$ to denote the set of primary edge subsets of size $r$, i.e.,
\begin{align*}
\mA_r=\big\{ A\subseteq E:\ A \text{ is primary and } |A|=r \big\}.\footnotemark
\end{align*}\footnotetext{We refer the reader to Example~\ref{eg-2} in Section~\ref{sec_algo2} for illustrations of $\mA_r$.} Guang and Yeung~\cite{GY-SNC-Reduction} showed that $\lvert \mA_r \rvert$ is an improved lower bound on the field size for the existence of SLNCs, and the improvement can be significant.

Now, we present the construction of SLNCs of Cai and Yeung~\cite{Cai-Yeung-SNC-IT}. Let $\w$ and $r$ be the information rate and the security level, respectively, and $n\triangleq \w+r\leq C_{\min}$. Let $\mC_n$ be an $n$-dimensional linear network code with global encoding kernels $\vf^{\,(n)}_e$, $e\in E$ over a finite field $\Fq$ on the network $G$. Let $\vb_1^{\,(n)}$, $\vb_2^{\,(n)}$, $\cdots$, $\vb_\w^{\,(n)}$ be $\w$ linearly independent column vectors in $\Fq^n$ such that
\begin{equation}\label{secure_condition}
\big\langle  \vb_i^{\,(n)}:\ 1\leq i \leq \w  \big\rangle \bigcap
\big\langle \vf^{\,(n)}_e:\ e\in A \big\rangle=\{\bzero\},\quad \forall~A\in \mA_r.
\end{equation}
Let $\vb_{\w+1}^{\,(n)}$, $\vb_{\w+2}^{\,(n)}$, $\cdots$, $\vb_n^{\,(n)}$ be another $n-\w$ column vectors in $\Fq^n$ such that the total $n$ vectors $\vb_1^{\,(n)}$, $\vb_2^{\,(n)}$, $\cdots$, $\vb_n^{\,(n)}$ are linearly independent, and let $Q^{(n)}=\left[ \vb_1^{\,(n)} \ \ \vb_2^{\,(n)} \ \  \cdots \ \ \vb_n^{\,(n)} \right]$. Then, $(Q^{(n)})^{-1}\cdot\,\mC_n$, the transformation of $\mC_n$ by $(Q^{(n)})^{-1}$ (cf.~Proposition~\ref{prop}), is an $n$-dimensional SLNC of information rate $\w$ and security level $r$. In using the SLNC $(Q^{(n)})^{-1}\cdot\,\mC_n$, we let $\mathbf{m}$, a row $\w$-vector in $\Fq^{\w}$, be the value of the source message $M$, and $\mathbf{k}$, a row $r$-vector in $\Fq^{r}$, be the value of the random key $K$. Then $\mathbf{x}=\big(\mathbf{m} \ \ \mathbf{k} \big)$ is the input of the source node $s$. With this setting, it was proved in~\cite{Cai-Yeung-SNC-IT} that this coding scheme not only multicasts the source message $M$ to all the sink nodes at rate $\w$ but also achieves security level~$r$.


\section{Secure Network Coding for a Fixed Rate and a Flexible Security Level}\label{Sec_v-s-l}

In this section, we consider the problem of designing a family of local-encoding-preserving SLNCs for a fixed information rate and multiple security levels. First, we present the following lemma that shows the existence of a family of local-encoding-preserving decodable linear network codes in which the linear network codes have distinct dimensions.

\begin{lemma}\label{thm_decoding_condition}
Let $\mC_{C_{\min}}=\big\{ \vf^{\,(C_{\min})}_e:~e\in E \big\}$ be a $C_{\min}$-dimensional decodable linear network code over the finite field $\Fq$ on the network $G=(V,E)$, where $C_{\min}=\min_{t\in T}C_t$. Let
\begin{align*}
\mC_{n}=\big[ I_{n} \ \ {\bf 0} \big] \cdot \mC_{C_{\min}} =\Big\{\vf_e^{\,(n)}\triangleq \big[ I_{n} \ \ {\bf 0} \big] \cdot \vf^{\,(C_{\min})}_e:~e\in E \Big\},\quad 1 \leq n \leq C_{\min},
\end{align*}
where we use ${\bf 0}$ in bold face to stand for an all-zero matrix of size $n\times (C_{\min}-n)$ and thus $\vf_e^{\,(n)}$ is the sub-vector of $\vf_e^{\,(C_{\min})}$ containing the first $n$ components.
Then, $\big\{ \mC_n:~n=1,2,\cdots,C_{\min} \big\}$ constitute a family of decodable linear network codes over $\Fq$ on $G$ with dimensions from 1 to $C_{\min}$, and have the same local encoding kernels at the non-source nodes.
\end{lemma}
\begin{IEEEproof}
This lemma can be proved straightforwardly by involving Corollary~2 in \cite{part1}, in which we take the column vector $\vec{\ell}$ to be the all-zero column vector $\bzero$ repeatedly with appropriate dimension.
\end{IEEEproof}

Based on the SLNC construction at the end of Section~\ref{Sec_Preli} and Lemma~\ref{thm_decoding_condition}, we naturally put forward the following approach for constructing a family of local-encoding-preserving SLNCs of the same information rate $\w$ and security levels from $0$ to $C_{\min}-\w$. First, we apply Lemma~\ref{thm_decoding_condition} to obtain a family of local-encoding-preserving linear network codes $\big\{ \mC_n:~n=\w, \w+1, \cdots ,C_{\min} \big\}$ of dimensions from $\w$ to $C_{\min}$. Next, for each $n$-dimensional linear network code $\mC_n$, we let $r=n-\w$ and design an $n\times n$ invertible matrix $Q^{(n)}$ satisfying \eqref{secure_condition}. Then, we construct a family of SLNCs $\big\{ (Q^{(n)})^{-1}\cdot\mC_n:~n=\w, \w+1, \cdots, C_{\min} \big\}$, which have the same rate $\w$ and different security levels from $0$ to $C_{\min}-\w$.

However, the above approach not only requires the construction of the matrix $Q^{(n)}$ for each $n$, incurring a high computational complexity, but also requires the source node $s$ to store all the matrices $Q^{(n)}$.\footnote{\label{footnote} The computational complexity of the construction of $Q^{(n)}$ is shown to be $\mO\big(\w n^3|\mA_{r}|+\w n |\mA_{r}|^2+r n^2 \big)$ in Appendix~A in~\cite{part1}, and the storage cost is $\mO\big( n^2 \big)$.} To avoid these shortcomings, we put forward the following more efficient approach to solve this problem.

\subsection{Approach and Technique}

We first present the following lemma which is instrumental to our approach.

\begin{lemma}\label{lemma_iff_cond}
Let $\w$ and $r$ be the information rate and the security level, respectively, and let $n=\w+r$. Consider an $n$-dimensional linear network code $\mC_{n}=\big\{\vf_e^{\,(n)}:~e\in E \big\}$ over a finite field $\Fq$ on the network $G$, and an $n\times n$
invertible matrix $Q^{(n)}=\Big[ \vb_1^{\,(n)}\ \ \vb_2^{\,(n)}\ \ \cdots \ \ \vb_{n}^{\,(n)} \Big]$ over $\Fq$. Then, $(Q^{(n)})^{-1}\cdot\,\mC_{n}$ is a rate-$\w$ and security-level-$r$ SLNC over $\Fq$ on $G$ if and only if \eqref{secure_condition} is satisfied, i.e.,
\begin{align*}
\big\langle  \vb_i^{\,(n)}:\ 1\leq i \leq \w  \big\rangle \bigcap
\big\langle \vf^{\,(n)}_e:\ e\in A \big\rangle=\{\bzero\},
\quad \forall~A\in \mA_r.
\end{align*}
\end{lemma}

In order to prove Lemma~\ref{lemma_iff_cond}, we need the following lemma. Before stating this lemma, we first let
\begin{align*}
\mB_j^{(n)}=\big\langle  \vb_i^{\,(n)}:~1\leq i \leq j \big\rangle \quad \text{and} \quad \mL_A^{(n)}=\big\langle  \vf_e^{\,(n)}:~e\in A \big\rangle
\end{align*}
for any $ j \leq n \leq C_{\min}$ and any $A\subseteq E$.

\begin{lemma}\label{lem_equi_Er_vs_Ar}
Let $\w$ and $r$ be the information rate and the security level, respectively, and let $n=\w+r$. Consider an $n$-dimensional linear network code $\mC_{n}=\big\{\vf_e^{\,(n)}:~e\in E \big\}$ over a finite field $\Fq$ on the network $G$ and $\w$ linearly independent column $n$-vectors $\vb_1^{\,(n)}$, $\vb_2^{\,(n)}$, $\cdots$, $\vb_{\w}^{\,(n)}$. Then,
\begin{align}\label{equ2-lem_equi_Er_vs_Ar}
\mB_\w^{(n)} \bigcap \mL_A^{(n)}=\{\bzero\},
\qquad \forall~A\in \mA_r
\end{align}
if and only if
\begin{align}\label{equ1-lem_equi_Er_vs_Ar}
\mB_\w^{(n)} \bigcap \mL_A^{(n)}=\{\bzero\},
\qquad \forall~A\in \mE_r,
\end{align}
where $\mE_r$ is the set of the edge subsets of size not larger than $r$, i.e., $\mE_r=\{ A\subseteq E:\ |A|\leq r \}$.
\end{lemma}
\begin{IEEEproof}
The ``if'' part is evident since $\mA_r \subseteq \mE_r$. We prove the ``only if'' part in the following.
Consider an arbitrary wiretap set $A$ in $\mE_r$ (not necessarily regular) and a minimum cut $\CUT_A$ between $s$ and $A$. Clearly, this minimum cut $\CUT_A$ is regular and $|\CUT_A|\leq r$. Since $r \leq n \leq C_{\min}$, we can choose another $r'\triangleq r-|\CUT_A|$ edges $e_1, e_2, \cdots, e_{r'}$ such that the edge subset $B \triangleq \CUT_A \bigcup \{e_1, e_2, \cdots, e_{r'}\}$ is regular. Let $B'$ be the primary minimum cut between $s$ and $B$, which, together with $|B|=r$, implies that $B'$ is primary and $|B'|=r$. We thus obtain that the wiretap set $A$ in $\mE_r$ is separated by the primary edge subset $B'$ in $\mA_r$ from $s$.
It follows from the mechanism of network coding that
\begin{align*}
\big\langle \vf^{\,(n)}_e:\ e\in A \big\rangle \subseteq \big\langle \vf^{\,(n)}_e:\ e\in B' \big\rangle,
\end{align*}
which, together with \eqref{equ2-lem_equi_Er_vs_Ar} implies that
$$\mB_\w^{(n)} \bigcap \mL_A^{(n)}=\{\bzero\}.$$
The lemma is proved.
\end{IEEEproof}

\begin{IEEEproof}[Proof of Lemma~\ref{lemma_iff_cond}]
The ``if'' part can be proved immediately by combining Lemma~\ref{lem_equi_Er_vs_Ar} with the proof of Theorem~2 in \cite{Cai-Yeung-SNC-IT} (cf.~\cite[Section~\Rmnum{5}]{Cai-Yeung-SNC-IT}). So, it suffices to prove the ``only if'' part. In fact, Remark~1 in \cite{Cai-Yeung-SNC-IT} gives an intuitive explanation about the ``only if'' part. In the following, we will give a rigorous proof.

Let $Q^{(n)}=\Big[ \vb_1^{\,(n)}\ \ \vb_2^{\,(n)}\ \ \cdots \ \ \vb_{n}^{\,(n)} \Big]$ be an $n \times n$ invertible matrix over $\Fq$ such that $(Q^{(n)})^{-1}\cdot\mC_{n}$ is a rate-$\w$ and security-level-$r$ SLNC over $\Fq$ on $G$. Assume the contrary that  there exists a nonzero column $n$-vector $\vv$ such that
$\vv \in \big\langle  \vb_i^{\,(n)}:\ 1\leq i \leq \w  \big\rangle \bigcap
\big\langle \vf^{\,(n)}_e:\ e\in A \big\rangle$
for some wiretap subset $A$ in $\mA_{r}$. Then there exist $\alpha_1$, $\alpha_2$, $\cdots$, $\alpha_\w$ in $\Fq$, not all zero, such that
\begin{align}\label{equ1_lemma_iff_cond}
\vv=\sum_{i=1}^\w\alpha_i\vb_i^{\,(n)},
\end{align}
and another $r$ elements in $\Fq$, denoted by $\beta_e$, $e\in A$, which are not all zero, such that
\begin{align}\label{equ2_lemma_iff_cond}
\vv=\sum_{e\in A} \beta_e\vf_e^{\,(n)}.
\end{align}
Combining \eqref{equ1_lemma_iff_cond} and \eqref{equ2_lemma_iff_cond}, we have
\begin{align}\label{lemma_iff_cond_equ1}
\vv=\sum_{e\in A} \beta_e\vf_e^{\,(n)}=\sum_{i=1}^\w\alpha_i\vb_i^{\,(n)}.
\end{align}
With the invertibility of the matrix $Q^{(n)}$, the equality \eqref{lemma_iff_cond_equ1} is equivalent to the following:
\begin{align}\label{lemma_iff_cond_equ2}
\bzero \neq \vv\,' \triangleq (Q^{(n)})^{-1}\cdot\vv=\sum_{e\in A} \beta_e \cdot (Q^{(n)})^{-1} \cdot  \vf_e^{\,(n)} =
\sum_{i=1}^\w\alpha_i \cdot(Q^{(n)})^{-1}\cdot \vb_i^{\,(n)}=\sum_{i=1}^\w\alpha_i \vec{1}^{\,(n)}_i,
\end{align}
where we note that $(Q^{(n)})^{-1}\cdot \vb_i^{\,(n)}=\vec{1}^{\,(n)}_i$, $1\leq i \leq \w$, with $\vec{1}^{\,(n)}_i$ being the $n$-column indicator vector of the $i$th component. Let $ \vv\,'=\big( v_1', v_2', \cdots, v_n' \big)$. By \eqref{lemma_iff_cond_equ2}, we obtain
\begin{align}\label{lemma_iff_cond_equ3}
v_{\w+1}'=v_{\w+2}'=\cdots=v_n'=0.
\end{align}

Now, we consider an arbitrary row $\w$-vector $\mathbf{m}$ in $\Fq^{\w}$ and an arbitrary row $r$-vector $\mathbf{k}$ in $\Fq^{r}$. Together with \eqref{lemma_iff_cond_equ3}, we have
\begin{align}
\big(\mathbf{m} \ \ \mathbf{k} \big) \cdot \vv\,' = \big(\mathbf{m} \ \ \mathbf{k} \big) \sum_{i=1}^\w\alpha_i \vec{1}^{\,(n)}_i = &\big(\mathbf{m} \ \ \mathbf{k} \big) \sum_{e\in A} \beta_e \cdot (Q^{(n)})^{-1} \cdot  \vf_e^{\,(n)}\nonumber\\
= &
\sum_{e\in A} \beta_e \cdot \Big[ \big(\mathbf{m} \ \ \mathbf{k} \big) \cdot (Q^{(n)})^{-1} \cdot  \vf_e^{\,(n)}\Big] =
\sum_{e\in A} \beta_e \cdot y_e,\label{lemma_iff_cond_equ4}
\end{align}
where $y_e$ in $\Fq$ is the message transmitted in the edge $e$ when the source message $M$ and the random key $K$ takes the values $\mathbf{m}$ and $\mathbf{k}$, respectively.

Combining \eqref{lemma_iff_cond_equ3} and \eqref{lemma_iff_cond_equ4}, we thus obtain
\begin{align}\label{equ5_lemma_iff_cond}
\big(\mathbf{m} \ \ \mathbf{k} \big) \cdot \vv\,' = \mathbf{m} \cdot \big( v_1' \ \ v_2' \ \ \cdots \ \ v_\w' \big)^\top = \sum_{e\in A} \beta_e \cdot y_e.
\end{align}
In other words, the effects of the random key $K$ on the messages transmitted on the edges in $A$ can be removed by taking a linear combination of these messages. Also, we see from the second equality above that for any given set of messages transmitted on the edges in $A$, the source message $\mathbf{m}$ is constrained.

Next, we calculate the conditional probability
${\rm Pr}\big( M= \mathbf{m} | Y_A=y_A)$, where $Y_A=(Y_e,~e\in A)$ and $y_A=(y_e,~e\in A)$. We first let $F_A=\Big[(Q^{(n)})^{-1} \cdot  \vf_e^{\,(n)}:~e\in A\Big]$, so that
\begin{align}\label{equ6_lemma_iff_cond}
\big(\mathbf{m} \ \ \mathbf{k} \big) \cdot F_A=y_A.
\end{align}
Then
\begin{align}
&{\rm Pr}\big( M= \mathbf{m} | Y_A=y_A)\nonumber\\
&=\frac{{\rm Pr}\big( M= \mathbf{m},~ Y_A=y_A)}{{\rm Pr}\big( Y_A=y_A)}\nonumber\\
&=\frac{\sum_{\mathbf{k}'\in\Fq^r}{\rm Pr}\big( M=\mathbf{m},~K=\mathbf{k}',~Y_A=y_A)}{{\rm Pr}\big( Y_A=y_A)}\nonumber\\
&=\frac{\sum_{\mathbf{k}'\in\Fq^r}{\rm Pr}\big(Y_A=y_A|M=\mathbf{m},~K=\mathbf{k}')\cdot{\rm Pr}\big(M=\mathbf{m},~K=\mathbf{k}')}{{\rm Pr}\big( Y_A=y_A)}\nonumber\\
&=\frac{\sum\limits_{\mathbf{k}'\in\Fq^r \text{ s.t. } (\mathbf{m} \ \mathbf{k}') \cdot F_A=y_A}1\cdot{\rm Pr}\big(M=\mathbf{m},~K=\mathbf{k}')}
{\sum\limits_{\substack{(\mathbf{m}'\ \mathbf{k}')\in\Fq^n \text{ s.t. }\\ (\mathbf{m}' \ \mathbf{k}') \cdot F_A=y_A}}{\rm Pr}\big(Y_A=y_A|M=\mathbf{m}',~K=\mathbf{k}')\cdot{\rm Pr}\big(M=\mathbf{m}',~K=\mathbf{k}')}\nonumber\\
&=\frac{\sum\limits_{\mathbf{k}'\in\Fq^r \text{ s.t. } (\mathbf{m} \ \mathbf{k}') \cdot F_A=y_A}{\rm Pr}\big(M=\mathbf{m},~K=\mathbf{k}')}
{\sum\limits_{\substack{(\mathbf{m}'\ \mathbf{k}')\in\Fq^n \text{ s.t. }\\ (\mathbf{m}' \ \mathbf{k}') \cdot F_A=y_A}} {\rm Pr}\big(M=\mathbf{m}',~K=\mathbf{k}')}\nonumber\\
&=\frac{\#\big\{\mathbf{k}'\in\Fq^r:~(\mathbf{m} \ \mathbf{k}') \cdot F_A=y_A\big\}}
{\#\big\{(\mathbf{m}'\ \mathbf{k}')\in\Fq^n:~(\mathbf{m}' \ \mathbf{k}') \cdot F_A=y_A \big\}},\label{equ7_lemma_iff_cond}
\end{align}
where we use ``$\#\{\cdot\}$'' to stand for the cardinality of the set, and the last equality \eqref{equ7_lemma_iff_cond} holds because $M$ and $K$ are independent and uniformly distributed over $\Fq^\w$ and $\Fq^r$, respectively.

We further write
\begin{align}\label{equ8_lemma_iff_cond}
F_A=\begin{bmatrix} F_{A,\w}\\ F_{A,r}\end{bmatrix},
\end{align}
where $F_{A,\w}$ is the sub-matrix containing the first $\w$ row vectors of $F_A$, and   $F_{A,r}$ is the one containing the remaining $r$ row vectors of $F_A$. With \eqref{equ8_lemma_iff_cond}, we have
\begin{align}\label{equ9_lemma_iff_cond}
\#\big\{\mathbf{k}'\in\Fq^r:~(\mathbf{m} \ \mathbf{k}') \cdot F_A=y_A\big\}
=\#\big\{\mathbf{k}'\in\Fq^r:~\mathbf{k}'F_{A,r}=y_A-\mathbf{m}F_{A,\w} \big\}=q^{r-\Rank(F_{A,r})},
\end{align}
where the last equality holds since there exists a solution $\mathbf{k}'$ (e.g., $\mathbf{k}$ by \eqref{equ6_lemma_iff_cond}) for the equation $\mathbf{k}'F_{A,r}=y_A-\mathbf{m}F_{A,\w}$.

Furthermore, since we have proved that the equality~\eqref{equ5_lemma_iff_cond} holds for all pairs $(\mathbf{m} \ \mathbf{k})$ of $\mathbf{m}\in \Fq^\w$ and $\mathbf{k}\in\Fq^r$ satisfying \eqref{equ6_lemma_iff_cond}, we obtain that
\begin{align}
&\#\big\{(\mathbf{m}'\ \mathbf{k}')\in\Fq^n:~(\mathbf{m}' \ \mathbf{k}') \cdot F_A=y_A \big\}\nonumber\\
&=\#\Big\{(\mathbf{m}'\ \mathbf{k}')\in\Fq^n:~(\mathbf{m}' \ \mathbf{k}') \cdot F_A=y_A \ \text{ and }\ \mathbf{m}' \cdot \big( v_1' \ v_2' \ \cdots \ v_\w' \big)^\top = \sum_{e\in A} \beta_e y_e \Big\}\nonumber\\
&=\#\bigcup_{\substack{\mathbf{m}'\in\Fq^{\w} \text{ s.t. }\\ \mathbf{m}' \cdot \big( v_1' \ v_2' \ \cdots \ v_\w' \big)^\top = \sum_{e\in A} \beta_e y_e}} \big\{(\mathbf{m}'\ \mathbf{k}')\in\Fq^n:~(\mathbf{m}' \ \mathbf{k}') \cdot F_A=y_A \big\}\nonumber\\
&=\sum_{\substack{\mathbf{m}'\in\Fq^{\w} \text{ s.t. }\\ \mathbf{m}' \cdot \big( v_1' \ v_2' \ \cdots \ v_\w' \big)^\top = \sum_{e\in A} \beta_e y_e}}
\#\big\{\mathbf{k}'\in\Fq^r:~(\mathbf{m}' \ \mathbf{k}') \cdot F_A=y_A \big\}\nonumber\\
&=\sum_{\substack{\mathbf{m}'\in\Fq^{\w} \text{ s.t. }\\ \mathbf{m}' \cdot \big( v_1' \ v_2' \ \cdots \ v_\w' \big)^\top = \sum_{e\in A} \beta_e y_e}}
\#\big\{\mathbf{k}'\in\Fq^r:~\mathbf{k}'F_{A,r}=y_A-\mathbf{m}'F_{A,\w}  \big\}\nonumber\\
&\leq \sum_{\substack{\mathbf{m}'\in\Fq^{\w} \text{ s.t. }\\ \mathbf{m}' \cdot \big( v_1' \ v_2' \ \cdots \ v_\w' \big)^\top = \sum_{e\in A} \beta_e y_e}}
q^{r-\Rank(F_{A,r})}\label{equ11_lemma_iff_cond}\\
&=\#\Big\{\mathbf{m}'\in\Fq^{\w}:~\mathbf{m}' \cdot \big( v_1' \ v_2' \ \cdots \ v_\w' \big)^\top = \sum_{e\in A} \beta_e y_e\Big\}\cdot q^{r-\Rank(F_{A,r})}\nonumber\\
&=q^{\w-1}\cdot q^{r-\Rank(F_{A,r})}\label{equ12_lemma_iff_cond},
\end{align}
where the inequality~\eqref{equ11_lemma_iff_cond} follows from the fact that for the given $\mathbf{m}'\in\Fq^\w$,
\begin{align*}
\#\big\{\mathbf{k}'\in\Fq^r:~\mathbf{k}'F_{A,r}=y_A-\mathbf{m}'F_{A,\w}  \big\}=\begin{cases}
q^{r-\Rank(F_{A,r})}, & \text{if $\mathbf{k}'F_{A,r}=y_A-\mathbf{m}'F_{A,\w}$ has a solution,}\\
0, & \text{otherwise.}
\end{cases}
\end{align*}
Substituting \eqref{equ9_lemma_iff_cond} and \eqref{equ12_lemma_iff_cond} into \eqref{equ7_lemma_iff_cond}, we immediately prove that
\begin{align*}
{\rm Pr}\big( M= \mathbf{m} | Y_A=y_A) \geq \frac{1}{q^{\w-1}}.
\end{align*}
Hence,
\begin{align*}
{\rm Pr}\big( M= \mathbf{m} | Y_A=y_A) \neq {\rm Pr}\big( M= \mathbf{m}\big)=\frac{1}{q^{\w}},
\end{align*}
a contradiction to $H(M)=H(M|Y_A)$, namely that the SLNC $(Q^{(n)})^{-1}\cdot\,\mC_{n}$ achieves the security level~$r$. The lemma is proved.
\end{IEEEproof}

Let $\mC_{C_{\min}}=\big\{ \vf^{\,(C_{\min})}_e:~e\in E \big\}$ be a $C_{\min}$-dimensional linear network code over $\Fq$ on the network $G$. Let $\w$ be the fixed information rate. By Lemma~\ref{thm_decoding_condition}, $$\mC_n=\Big\{\vf_e^{\,(n)}\triangleq \big[ I_{n} \ \ {\bf 0} \big] \cdot \vf^{\,(C_{\min})}_e:~e\in E \Big\},\footnotemark\quad n=\w, \w+1, \cdots, C_{\min}$$\footnotetext{Here, ${\bf 0}$ in bold face stands for an all-zero matrix of size $n\times (C_{\min}-n)$.}constitute a family of local-encoding-preserving linear network codes with dimensions from $\w$ to $C_{\min}$. Note that $C_{\w}$ can be regarded as an $\w$-dimensional SLNC with rate $\w$ and security level $0$. Further, for any $\w \times \w$ invertible matrix $Q^{(\w)}=\begin{bmatrix} \vb_1^{\,(\w)} & \vb_2^{\,(\w)} & \cdots & \vb_{\w}^{\,(\w)}\end{bmatrix}$, it follows from Proposition~\ref{prop} that $(Q^{(\w)})^{-1}\cdot\,\mC_\w$ also is an $\w$-dimensional SLNC with rate $\w$ and security level~$0$.

Now, consider any $n$-dimensional ($\w \leq n \leq C_{\min}-1$) SLNC $(Q^{(n)})^{-1}\cdot\,\mC_n$ with the fixed rate $\w$ and security level $r\triangleq n-\w$, where we write $Q^{(n)}=\begin{bmatrix} \vb_1^{\,(n)} & \vb_2^{\,(n)} & \cdots & \vb_{n}^{\,(n)}\end{bmatrix}$. By Lemma~\ref{lemma_iff_cond}, this is equivalent to that $Q^{(n)}$ is an $n\times n$ invertible matrix satisfying
\begin{align*}
\big\langle  \vb_i^{\,(n)}:\ 1\leq i \leq \w  \big\rangle \bigcap
\big\langle \vf^{\,(n)}_e:\ e\in A \big\rangle=\{\bzero\},
\quad \forall~A\in \mA_r.
\end{align*}
Based on the SLNC $(Q^{(n)})^{-1}\cdot\,\mC_n$, we will construct an $(n+1)$-dimensional SLNC $(Q^{(n+1)})^{-1}\cdot\,\mC_{n+1}$ with rate $\w$ and security level $r+1$. Our idea is to design an appropriate row $\w$-vector $\vc=\big(c_1 \ \  c_2 \ \ \cdots \ \ c_\w \big)\in \Fq^{\w}$ to obtain $\w$ column $(n+1)$-vectors $\vb_i^{\,(n+1)}\triangleq \begin{bmatrix} \vb_i^{\,(n)}\\ c_i  \end{bmatrix}$, $1\leq i \leq \w$, such that $\vb_i^{\,(n+1)}$, $1\leq i \leq \w$, are linearly independent, and
\begin{align*}
\big\langle  \vb_i^{\,(n+1)}:\ 1\leq i \leq \w  \big\rangle \bigcap
\big\langle \vf^{\,(n+1)}_e:\ e\in A \big\rangle=\{\bzero\},
\quad \forall~A\in \mA_{r+1}.
\end{align*}
Then with $\vb_i^{\,(n+1)}$, $1\leq i \leq \w$, we construct another $n+1-\w$ column $(n+1)$-vectors $\vb_{\w+1}^{\,(n+1)}$, $\vb_{\w+2}^{\,(n+1)}$, $\cdots$, $\vb_{n+1}^{\,(n+1)}$ such that $\Big[ \vb_{1}^{\,(n+1)} \ \ \vb_{2}^{\,(n+1)} \ \ \cdots\ \ \vb_{n+1}^{\,(n+1)} \Big]$, denoted by $Q^{(n+1)}$, is an $(n+1)\times (n+1)$ invertible matrix. In this way, we can construct an $(n+1)$-dimensional SLNC $(Q^{(n+1)})^{-1}\cdot\mC_{n+1}$ that retains the fixed rate $\w$, achieves a higher security level $r+1$, and has the same local encoding kernels as $(Q^{(n)})^{-1}\cdot\mC_{n}$ at all the non-source nodes by Proposition~\ref{prop}.

Next, we will design such an appropriate vector $\vc$, which is the crucial to our proposed approach. Based on $\mC_{n}$, we partition $\mA_{r+1}$ into two disjoint subsets:
\begin{align}
\widetilde{\mA}_{r+1}' & = \big\{ A\in \mA_{r+1}:~\vf_e^{\,(n)},~e\in A,~\text{are linearly dependent} \big\},\label{A_r+1'}\\
\widetilde{\mA}_{r+1}'' & =\big\{ A\in \mA_{r+1}:~\vf_e^{\,(n)},~e\in A,~\text{are linearly independent} \big\}.\label{A_r+1''}
\end{align}

\begin{lemma}\label{lem7}
For $\widetilde{\mA}_{r+1}'$ and $\widetilde{\mA}_{r+1}''$,
\begin{align}\label{A_r+1'_alt_form}
\widetilde{\mA}_{r+1}' = \big\{ A\in \mA_{r+1}:~\mB_{\w}^{(n)} \bigcap \mL_A^{(n)}=\{ \bzero \} \big\},
\end{align}
and
\begin{align}\label{A_r+1''_alt_form}
\widetilde{\mA}_{r+1}'' & =\big\{A\in \mA_{r+1}:~\mB_{\w}^{(n)} \bigcap \mL_A^{(n)}\neq \{ \bzero \} \big\}.
\end{align}
\end{lemma}

\begin{IEEEproof}
It suffices to prove that for any $A\in \mA_{r+1}$, the vectors $\vf_e^{\,(n)}$, $e\in A$, are linearly independent if and only if $\mB_\w^{(n)}\bigcap\mL_A^{(n)}\neq \{\bzero\}$. For the ``only if'' part, since $\vf_e^{\,(n)}$, $e\in A$, are linearly independent, we have $\dim\big(\mL_A^{(n)}\big)=r+1$. Thus,
\begin{align*}
\dim\big(\mB_\w^{(n)}\bigcap\mL_A^{(n)} \big)=
\dim\big(\mB_\w^{(n)}\big)+\dim\big(\mL_A^{(n)}\big)
-\dim\big(\mB_\w^{(n)}+\mL_A^{(n)}\big)
\geq \w+(r+1)-n=1,
\end{align*}
where the inequality follows from $\dim\big(\mB_\w^{(n)}+\mL_A^{(n)}\big)\leq n=\w+r$. This implies $\mB_\w^{(n)}\bigcap\mL_A^{(n)}\neq \{\bzero\}$.

For the ``if'' part, we let $A'\subseteq A$ such that $\big\{ \vf_e^{\,(n)}:~e\in A' \big\}$ is a maximal linearly independent subset of $\big\{ \vf_e^{\,(n)}:~e\in A \big\}$. Then, we have
\begin{align}\label{equ_thm_frvs_case2}
\mB_\w^{(n)}\bigcap\mL_{A'}^{(n)}=\mB_\w^{(n)}\bigcap\mL_A^{(n)}\neq \{\bzero\}.
\end{align}
This immediately implies $A'=A$, because otherwise $|A'|\leq r$, which together with $\mB_\w^{(n)}\bigcap\mL_{A'}^{(n)}\neq \{\bzero\}$ in~\eqref{equ_thm_frvs_case2}, contradicts to \eqref{equ1-lem_equi_Er_vs_Ar} and thus \eqref{equ2-lem_equi_Er_vs_Ar} by Lemma~\ref{lem_equi_Er_vs_Ar}. Therefore, $\vf_e^{\,(n)}$, $e\in A$ are linearly independent. The lemma is proved.
\end{IEEEproof}

We first consider those wiretap sets in $\widetilde{\mA}_{r+1}'$. The following theorem asserts that any vector $\vc\in \Fq^\w$ is feasible for the wiretap sets in $\widetilde{\mA}_{r+1}'$.

\begin{thm}\label{thm_frvs_case1}
For any row $\w$-vector $\vc=( c_1 \ \ c_2 \ \ \cdots \ \ c_\w )\in \Fq^{\w}$, the following are satisfied:
\begin{itemize}
  \item the column $(n+1)$-vectors $\vb_i^{\,(n+1)}=\begin{bmatrix} \vb_i^{\,(n)}\\ c_i  \end{bmatrix}$, $1\leq i \leq \w$, are linearly independent;
  \item $\mB_\w^{(n+1)}\bigcap\mL_A^{(n+1)}=\{\bzero\}$, $\forall~ A\in \widetilde{\mA}_{r+1}'$.
\end{itemize}
\end{thm}
\begin{IEEEproof}
First, we note that the column $n$-vectors $\vb_i^{\,(n)}$, $1\leq i \leq \w$, are linearly independent, and so are the column $(n+1)$-vectors $\vb_i^{\,(n+1)}=\begin{bmatrix} \vb_i^{\,(n)}\\ c_i \end{bmatrix}$, $1\leq i \leq \w$.

Next, we prove by contradiction that for any $\vc=( c_1 \ \ c_2 \ \ \cdots \ \ c_\w )\in \Fq^{\w}$,
$$\mB_\w^{(n+1)}\bigcap\mL_A^{(n+1)}=\{\bzero\},\quad \forall~ A\in \widetilde{\mA}_{r+1}'.$$ Assume that there exists an edge subset $A\in \widetilde{\mA}_{r+1}'$ such that $\mB_\w^{(n+1)}\bigcap\mL_A^{(n+1)} \neq \{\bzero\}$, and let $\vv^{\,(n+1)}=\big[ v_1\ \ v_2\ \ \cdots \ \ v_{n+1} \big]^\top$ be a nonzero vector in $\mB_\w^{(n+1)}\bigcap\mL_A^{(n+1)}$. Then, there exist $\alpha_1$, $\alpha_2$, $\cdots$, $\alpha_\w$ in $\Fq$, not all zero, such that
\begin{align}\label{equ1'_thm_frvs_case1}
\vv^{\,(n+1)}=\sum_{i=1}^\w\alpha_i\vb_i^{\,(n+1)},
\end{align}
and another $r+1$ elements in $\Fq$, denoted by $\beta_e$, $e\in A$, which are not all zero, such that
\begin{align}\label{equ2'_thm_frvs_case1}
\vv^{\,(n+1)}=\sum_{e\in A} \beta_e\vf_e^{\,(n+1)}.
\end{align}
We further write \eqref{equ1'_thm_frvs_case1} and \eqref{equ2'_thm_frvs_case1} respectively as
\begin{align}\label{equ1_thm_frvs_case1}
\begin{bmatrix} \vv^{\,(n)}\\ v_{n+1}\end{bmatrix}
=\sum_{i=1}^\w \alpha_i \begin{bmatrix}\vb_i^{\,(n)}\\c_{i}\end{bmatrix},
\end{align}
and
\begin{align}\label{equ2_thm_frvs_case1}
\begin{bmatrix} \vv^{\,(n)}\\ v_{n+1}\end{bmatrix}
=\sum_{e\in A}\beta_e\begin{bmatrix} \vf_e^{\,(n)}\\ f_{e,n+1}\end{bmatrix},
\end{align}
where $\vv^{\,(n)}$ is the sub-vector of $\vv^{\,(n+1)}$ obtained by deleting the last component $v_{n+1}$, i.e., $\vv^{\,(n)}=\big[ v_1\ \ v_2\ \ \cdots \ \ v_{n} \big]^\top$, and $f_{e,n+1}$ is the last component of $\vf_e^{\,(n+1)}$. Combining \eqref{equ1_thm_frvs_case1} and \eqref{equ2_thm_frvs_case1}, we immediately obtain
\begin{align*}
\vv^{\,(n)}=\sum_{i=1}^\w \alpha_i\vb_i^{\,(n)}
=\sum_{e\in A} \beta_e\vf_e^{\,(n)},
\end{align*}
which implies $\vv^{\,(n)}\in \mB_\w^{(n)}\bigcap \mL_A^{(n)}$. On the other hand, since $\vb_1^{\,(n)}$, $\vb_2^{\,(n)}$, $\cdots$, $\vb_\w^{\,(n)}$ are linearly independent, and $\alpha_1$, $\alpha_2$, $\cdots$, $\alpha_\w$ are not all zero, we immediately have $\vv^{\,(n)}\neq \bzero$, implying that  $\mB_\w^{(n)}\bigcap\mL_A^{(n)} \neq \{\bzero\}$. This is a contradiction to the assumption that $A\in \widetilde{\mA}_{r+1}'$ (cf.~\eqref{A_r+1'_alt_form} in Lemma~\ref{lem7}). The theorem is proved.
\end{IEEEproof}

Next, we consider the vectors $\vc \in\Fq^\w$ that are feasible for the wiretap sets in $\widetilde{\mA}_{r+1}''$. We first prove the following lemma.

\begin{lemma}\label{lem_frvs_2}
For any edge subset $A\in \widetilde{\mA}_{r+1}''$, $\dim\big(\mB_\w^{(n)}\bigcap\mL_A^{(n)}\big)=1$.
\end{lemma}
\begin{IEEEproof}
Let $A$ be an arbitrary edge subset in $\widetilde{\mA}_{r+1}''$. By \eqref{A_r+1''_alt_form} in Lemma~\ref{lem7}, we assume that there exist two linearly independent vectors $\vv_1^{\,(n)}$ and $\vv_2^{\,(n)}$ in $\mB_\w^{(n)}\bigcap\mL_A^{(n)}$. Then, there exist $(\alpha_{1,i},~1\leq i \leq \w)$ and $(\alpha_{2,i},~1\leq i \leq \w)$  in $\Fq^{\w}$, and  $(\beta_{1,e},~e\in A)$ and $(\beta_{2,e},~e\in A)$ in $\Fq^{r+1}$ such that
\begin{align}\label{v_1}
\vv_1^{\,(n)}=\sum_{i=1}^\w\alpha_{1,i}\vb_i^{\,(n)}
=\sum_{e\in A}\beta_{1,e}\vf_e^{\,(n)},
\end{align}
and
\begin{align}\label{v_2}
\vv_2^{\,(n)}=\sum_{i=1}^\w\alpha_{2,i}\vb_i^{\,(n)}
=\sum_{e\in A}\beta_{2,e}\vf_e^{\,(n)}.
\end{align}

The linear independence of $\vv_1^{\,(n)}$ and $\vv_2^{\,(n)}$ implies that both of them are nonzero. Together with the linear independence of $\vb_1^{\,(n)}$, $\vb_2^{\,(n)}$, $\cdots$, $\vb_\w^{\,(n)}$, we obtain that $(\alpha_{1,i},~1\leq i \leq \w)$ and $(\alpha_{2,i},~1\leq i \leq \w)$ are nonzero vectors.

Next, we prove that $\beta_{1,e}$, $e\in A$ are all nonzero. Assume otherwise, i.e.,  $\beta_{1,e'}=0$ for some $e'\in A$. Then, by \eqref{v_1} we have
\begin{align}\label{equ1_lem_frvs_2}
\bzero\neq \vv_1^{\,(n)}\in \mB_\w^{(n)}\bigcap \big\langle \vf_e^{\,(n)}:~e \in A\setminus \{e'\} \big\rangle=\mB_\w^{(n)}\bigcap\mL_{A\setminus \{e'\}}^{(n)}.
\end{align}
We further note that the edge subset $A\setminus \{e'\}$ has cardinality $r$, implying that $A\setminus \{e'\}\in \mE_r$. This is a contradiction to $\mB_\w^{(n)}\bigcap\mL_{B}^{(n)}= \{\bzero\}$ for all $B\in\mE_r$, which follows from Lemmas~\ref{lemma_iff_cond} and \ref{lem_equi_Er_vs_Ar} because $(Q^{(n)})^{-1}\cdot\mC_n$ is an $n$-dimensional SLNC with rate $\w$ and security level $r$. Thus, $\beta_{1,e}$, $e\in A$ are all nonzero. Likewise, $\beta_{2,e}$, $e\in A$ are all nonzero.

Fix any $e''\in A$. Then $\beta_{1,e''}$ and $\beta_{2,e''}$ are both nonzero. Let $\gamma=-\beta_{1,e''}/\beta_{2,e''}$ which is a nonzero element in $\Fq$. Then
\begin{align}\label{frvs_2}
\beta_{1,e''}+\gamma\beta_{2,e''}=0.
\end{align}
On the other hand, $\vv_1^{\,(n)}+\gamma\vv_2^{\,(n)}\neq \bzero$ because $\vv_1^{\,(n)}$ and $\vv_2^{\,(n)}$ are linearly independent. By \eqref{v_1} and \eqref{v_2}, we obtain
\begin{align*}
\vv_1^{\,(n)}+\gamma\vv_2^{\,(n)}&=\sum_{i=1}^{\w}(\alpha_{1,i}
+\gamma\alpha_{2,i})\vb_i^{\,(n)}\\
&=\sum_{e\in A}(\beta_{1,e}+\gamma\beta_{2,e})\vf_e^{\,(n)}\\
&=\sum_{e\in A\setminus\{e''\}} (\beta_{1,e}+\gamma\beta_{2,e})\vf_e^{\,(n)},
\end{align*}
where the last equality follows from \eqref{frvs_2}.

Therefore, the nonzero vector $\vv_1^{\,(n)}+\gamma\vv_2^{\,(n)}$ is in $\mB_\w^{(n)}\bigcap \big\langle \vf_e^{\,(n)}:~e\in A\setminus\{e''\} \big\rangle=\mB_\w^{(n)}\bigcap\mL_{A\setminus \{e''\}}^{(n)}$, which implies
\begin{align*}
\mB_\w^{(n)}\bigcap\mL_{A\setminus \{e''\}}^{(n)} \neq \{\bzero\}.
\end{align*}
By the same argument following \eqref{equ1_lem_frvs_2}, this contradicts $\mB_\w^{(n)}\bigcap\mL_{B}^{(n)}= \{\bzero\}$ for all $B\in\mE_r$. Thus we have proved that there cannot exist $\vv_1^{\,(n)}$ and $\vv_2^{\,(n)}$ in $\mB_\w^{(n)}\bigcap\mL_A^{(n)}$ that are linearly independent. Together with \eqref{A_r+1''_alt_form} in Lemma~\ref{lem7}, this implies that $\dim\big(\mB_\w^{(n)}\bigcap\mL_A^{(n)}\big)=1$. The lemma is proved.
\end{IEEEproof}

\begin{thm}\label{thm_frvs_case2}
For each wiretap set $A\in \widetilde{\mA}_{r+1}''$, define a set of row $\w$-vectors as follows:
\begin{align}\label{Gamma_A}
\begin{split}
\Gamma_A=\Big\{ \vc=\big( c_1 \ \ c_2 \ \ \cdots \ \ c_\w \big)\in\Fq^{\w}:~\sum_{i=1}^{\w}\alpha_i c_i=\sum_{e\in A}\beta_e f_{e,n+1},
 \text{ where }  \big( \alpha_i,~1\leq i \leq \w \big)\in \Fq^{\w}\\
 \text{ and } \big( \beta_e,~e\in A \big)\in \Fq^{r+1}  \text{ s.t. } \sum_{i=1}^{\w}\alpha_i \vb_i^{\,(n)}=\sum_{e\in A} \beta_e \vf_e^{\,(n)}\neq \bzero \Big\},
\end{split}
\end{align}
where $f_{e,n+1}$ is the last component of the global encoding kernel $\vf_e^{\,(n+1)}$. Then for any row $\w$-vector $\vc=\big( c_1 \ \ c_2 \ \ \cdots \ \ c_\w \big) \in \Fq^\w \setminus \bigcup_{A\in \widetilde{\mA}_{r+1}''} \Gamma_A$,
the following are satisfied:
\begin{itemize}
  \item the column $(n+1)$-vectors $\vb_i^{\,(n+1)}=\begin{bmatrix} \vb_i^{\,(n)}\\ c_i  \end{bmatrix}$, $1\leq i \leq \w$, are linearly independent;
  \item $\mB_\w^{(n+1)}\bigcap\mL_A^{(n+1)}=\{\bzero\}$, $\forall~ A\in \widetilde{\mA}_{r+1}''$.
\end{itemize}
\end{thm}
\begin{rem}
In the definition of $\Gamma_A$ in~\eqref{Gamma_A}, by virtue of Lemma~\ref{lem_frvs_2}, there always exist $\big( \alpha_i,~1\leq i \leq \w \big)\in \Fq^{\w}$ and $\big( \beta_e,~e\in A \big)\in \Fq^{r+1}$ that satisfy the required condition. Then for any given such $\big( \alpha_i,~1\leq i \leq \w \big)$ and $\big( \beta_e,~e\in A \big)$, there always exists $\vc\in\Fq^{\w}$ satisfying $\sum_{i=1}^{\w}\alpha_i c_i=\sum_{e\in A}\beta_e f_{e,n+1}$. Therefore, $\Gamma_A$ is always nonempty.
\end{rem}
\begin{IEEEproof}[Proof of Theorem~\ref{thm_frvs_case2}]
First, the $\w$ column $n$-vectors $\vb_i^{\,(n)}$, $1 \leq i \leq \w$, are linearly independent, and so are the $\w$ column $(n+1)$-vectors $\vb_i^{\,(n+1)}=\begin{bmatrix} \vb_i^{\,(n)}\\ c_i  \end{bmatrix}$, $1\leq i \leq \w$, for any row $\w$-vector $\vc=\big( c_1 \ \ c_2 \ \ \cdots \ \ c_\w \big)$ in $\Fq^{\w}$. Thus, we obtain
\begin{align}\label{equ4}
\dim\big(\mB_\w^{(n)}\big)=\dim\big(\mB_\w^{(n+1)}\big).
\end{align}

Let $A$ be an arbitrary wiretap set in $\widetilde{\mA}_{r+1}''$. By the definition (cf.~\eqref{A_r+1''}), we have $\dim\big(\mL_A^{(n)}\big)=|A|$. This implies
\begin{align}\label{equ5}
\dim\big(\mL_A^{(n+1)}\big)=\dim\big(\mL_A^{(n)}\big)=r+1,
\end{align}
or equivalently, $\vf_e^{\,(n+1)}$, $e\in A$ are linearly independent.
By \eqref{equ4} and \eqref{equ5}, we see that
\begin{align}\label{equ4_thm_frvs_case2}
\dim\big(\mB_\w^{(n+1)}\bigcap\mL_A^{(n+1)}\big)&=\dim\big(\mB_\w^{(n+1)}\big)
+\dim\big(\mL_A^{(n+1)}\big)-\dim\big(\mB_\w^{(n+1)}+\mL_A^{(n+1)}\big)\nonumber\\
&\leq \dim\big(\mB_\w^{(n)}\big)
+\dim\big(\mL_A^{(n)}\big)-\dim\big(\mB_\w^{(n)}+\mL_A^{(n)}\big)\nonumber\\
&=\dim\big(\mB_\w^{(n)}\bigcap\mL_A^{(n)}\big)= 1,
\end{align}
where the last equality follows from Lemma~\ref{lem_frvs_2}.

Let $\vv^{\,(n)}$ be any nonzero column $n$-vector in $\mB_\w^{(n)}\bigcap\mL_{A}^{(n)}$. It then follows from Lemma~\ref{lem_frvs_2} that \begin{align}\label{equ5_thm_frvs_case2}
\mB_\w^{(n)}\bigcap\mL_{A}^{(n)}=\big\{ \gamma\vv^{\,(n)},~\gamma\in \Fq \big\}.
\end{align}
Next, we will prove the claim that  $\dim\big(\mB_\w^{(n+1)}\bigcap\mL_A^{(n+1)}\big)=1$ if and only if the vector $\vc=\big( c_1 \ \ c_2 \ \ \cdots \ \ c_\w \big)$ satisfies
\begin{align}\label{frvs_5}
\sum_{i=1}^{\w}\alpha_i c_i =\sum_{e\in A}\beta_e f_{e,n+1},
\end{align}
where $\big(\alpha_i,~1\leq i \leq \w \big)$ and $\big( \beta_e,~e\in A \big)$ are two nonzero row vectors over $\Fq$ such that
\begin{align}\label{equ3_thm_frvs_case2}
\vv^{\,(n)}=\sum_{i=1}^{\w}\alpha_i\vb_i^{\,(n)}=\sum_{e\in A}\beta_e \vf_e^{\,(n)}.
\end{align}

For the ``if'' part, since $\vb_i^{\,(n)}$, $1 \leq i \leq \w$, are linearly independent and $\vf_e^{\,(n)}$, $e\in A$ are also linearly independent (because $A\in \widetilde{\mA}_{r+1}''$), it follows from \eqref{equ3_thm_frvs_case2} that the vectors $\big(\alpha_i,~1\leq i \leq \w \big)$ and $\big( \beta_e,~e\in A \big)$ are unique. Let $v \in \Fq$ be the common value of both sides of \eqref{frvs_5}. Then by combining \eqref{equ3_thm_frvs_case2} and \eqref{frvs_5}, we have
\begin{align*}
\bzero \neq \begin{bmatrix} \vv^{\,(n)}\\ v\end{bmatrix}
=\sum_{i=1}^\w \alpha_i \begin{bmatrix}\vb_i^{\,(n)}\\c_{i}\end{bmatrix}=
\sum_{e\in A}\beta_e\begin{bmatrix} \vf_e^{\,(n)}\\ f_{e,n+1}\end{bmatrix},
\end{align*}
which immediately implies $\dim\big(\mB_\w^{(n+1)}\bigcap\mL_A^{(n+1)}\big)\geq 1$. Together with \eqref{equ4_thm_frvs_case2}, it follows that $$\dim\big(\mB_\w^{(n+1)}\bigcap\mL_A^{(n+1)}\big)=1.$$

For the ``only if'' part, since $\dim\big(\mB_\w^{(n+1)}\bigcap\mL_A^{(n+1)}\big)=1$,
we let
$$\bzero\neq\vu^{\,(n+1)}\triangleq \begin{bmatrix} \vu^{\,(n)}\\ u_{n+1} \end{bmatrix}\in \mB_\w^{(n+1)}\bigcap\mL_A^{(n+1)},$$
where $\vu^{\,(n)}$ is a column $n$-vector and $u_{n+1}$ is an element in $\Fq$. Then, there exist unique vectors $\big(\alpha_i',~ 1 \leq i \leq \w \big)$ and $\big( \beta_e',~e\in A \big)$, both nonzero, such that
\begin{align*}
\vu^{\,(n+1)}=\sum_{i=1}^{\w}\alpha_i'\vb_i^{\,(n+1)}
=\sum_{e\in A}\beta_e' \vf_e^{\,(n+1)},
\end{align*}
or equivalently,
\begin{align}
\vu^{\,(n)}=\sum_{i=1}^{\w}\alpha_i'\vb_i^{\,(n)}
=\sum_{e\in A}\beta_e' \vf_e^{\,(n)},\label{equ1_thm_frvs_case2}
\end{align}
and
\begin{align}
u_{n+1}=\sum_{i=1}^{\w}\alpha_i' c_i
=\sum_{e\in A}\beta_e' f_{e, n+1}.\label{equ2_thm_frvs_case2}
\end{align}
From \eqref{equ1_thm_frvs_case2}, since $\vb_i^{\,(n)}$, $1 \leq i \leq \w$ is linearly independent and $\big(\alpha_i',~1\leq i \leq \w \big)$ is a nonzero vector, it follows that $\vu^{\,(n)}\neq \bzero$ and $\vu^{\,(n)}\in \mB_\w^{(n)} \bigcap\mL_A^{(n)}$. Thus, we have $\vu^{\,(n)}=\gamma\vv^{\,(n)}$ for some $\gamma\in \Fq\setminus\{0\}$. Together with \eqref{equ3_thm_frvs_case2}, we further have
\begin{align}\label{equ6_thm_frvs_case2}
\vu^{\,(n)}=\gamma\vv^{\,(n)}
=\sum_{i=1}^{\w}\gamma\alpha_i\vb_i^{\,(n)}
=\sum_{e\in A}\gamma\beta_e \vf_e^{\,(n)}.
\end{align}
Upon comparing \eqref{equ1_thm_frvs_case2} and \eqref{equ6_thm_frvs_case2}, we obtain
\begin{align}\label{equ7_thm_frvs_case2}
\big(\alpha_i',~ 1\leq i \leq \w \big)
=\gamma\big(\alpha_i,~1\leq i \leq \w \big)
\end{align}
and
\begin{align}\label{equ8_thm_frvs_case2}
\big( \beta_e',~e\in A \big)
=\gamma\big( \beta_e,~e\in A \big).
\end{align}
Thus, from \eqref{equ7_thm_frvs_case2}, \eqref{equ8_thm_frvs_case2}, and the second equality in \eqref{equ2_thm_frvs_case2}, we obtain \eqref{frvs_5}, proving the ``only if'' part of the claim.

Next, we define a set of row $\w$-vectors associated with each nonzero vector $\vv^{\,(n)}$ in $\mB_\w^{(n)}\bigcap\mL_A^{(n)}$ as follows:
\begin{align}\label{Gamma_v^n}
\Gamma_A\big(\vv^{\,(n)}\big)
=\Big\{ \vc=&\big( c_1 \ \ c_2 \ \ \cdots \ \ c_\w \big)\in\Fq^{\w}:~\sum_{i=1}^{\w}\alpha_i c_i=\sum_{e\in A}\beta_e f_{e,n+1},
 \text{ where }  \big( \alpha_i,~1\leq i \leq \w \big)\in \Fq^{\w}\nonumber\\
 &\text{ and } \big( \beta_e,~ e\in A \big)\in \Fq^{r+1}  \text{ s.t. } \vv^{\,(n)}=\sum_{i=1}^{\w}\alpha_i \vb_i^{\,(n)}=\sum_{e\in A} \beta_e \vf_e^{\,(n)}\neq \bzero \Big\}.
\end{align}
Then, in light of \eqref{equ4_thm_frvs_case2}, the foregoing claim is equivalent to \begin{align}\label{equ12_thm_frvs_case2}
\mB_\w^{(n+1)}\bigcap\mL_{A}^{(n+1)}\neq \{\bzero\}~\Longleftrightarrow~\vc\in \Gamma_A\big(\vv^{\,(n)}\big).
\end{align}
Since the LHS above does not depend on $\vv^{\,(n)}$, we see immediately that $\Gamma_A\big(\vv^{\,(n)}\big)$ also does not depend on $\vv^{\,(n)}$. Upon noting by \eqref{Gamma_A} that
\begin{align}\label{equ13_thm_frvs_case2}
\Gamma_A=\bigcup_{\vv^{\,(n)}\in \mB_\w^{(n)}\bigcap\mL_{A}^{(n)}\setminus \{\bzero\}} \Gamma_A\big(\vv^{\,(n)}\big),
\end{align}
we obtain that
\begin{align}\label{equ14_thm_frvs_case2}
\Gamma_A=\Gamma_A\big(\vv^{\,(n)}\big),
\end{align}
where $\vv^{\,(n)}$ is any vector in $\mB_\w^{(n)}\bigcap\mL_{A}^{(n)}\setminus \{\bzero\}$. Hence, \eqref{equ12_thm_frvs_case2} is equivalent to
\begin{align*}
\mB_\w^{(n+1)}\bigcap\mL_{A}^{(n+1)}\neq \{\bzero\}~\Longleftrightarrow~\vc\in \Gamma_A,
\end{align*}
or equivalently,
\begin{align*}
\mB_\w^{(n+1)}\bigcap\mL_{A}^{(n+1)} = \{\bzero\}~\Longleftrightarrow~\vc\in \Fq^\w\setminus \Gamma_A.
\end{align*}
Based on the above, the theorem can be proved immediately by considering all wiretap sets $A\in \widetilde{\mA}_{r+1}''$.
\end{IEEEproof}

By combining Theorems~\ref{thm_frvs_case1} and \ref{thm_frvs_case2}, we present the following theorem which gives the prescription for designing the vector $\vc$ for $\mA_{r+1}$.

\begin{thm}\label{thm_11}
Let $\Fq$ be a finite field of order $q>|\widetilde{\mA}_{r+1}''|$. Then the set $\Fq^\w \setminus \bigcup_{A\in \widetilde{\mA}_{r+1}''}\Gamma_A$ is nonempty, and for any $\vc=\big(c_1 \ \ c_2 \ \ \cdots \ \ c_\w \big)$ in this set, the following are satisfied:
\begin{itemize}
  \item the column $(n+1)$-vectors $\vb_i^{\,(n+1)}=\begin{bmatrix} \vb_i^{\,(n)}\\ c_i  \end{bmatrix}$, $1\leq i \leq \w$, are linearly independent;
  \item $\mB_\w^{(n+1)}\bigcap\mL_A^{(n+1)}=\{\bzero\}$, $\forall~ A\in \mA_{r+1}$.
\end{itemize}
\end{thm}

\begin{IEEEproof}
We only need prove that if $q>|\widetilde{\mA}_{r+1}''|$, then $\Fq^\w \setminus \bigcup_{A\in \widetilde{\mA}_{r+1}''}\Gamma_A$ is nonempty. The rest of the theorem follows immediately from Theorems~\ref{thm_frvs_case1}~and~\ref{thm_frvs_case2}.

Let $A$ be an arbitrary wiretap set in $\widetilde{\mA}_{r+1}''$. By the proof of Theorem~\ref{thm_frvs_case2}, we have $\Gamma_A=\Gamma_A\big( \vv^{\,(n)}\big)$ (cf.~\eqref{equ14_thm_frvs_case2}) for any nonzero vector $\vv^{\,(n)}\in \mB_\w^{(n)}\cap\mL_{A}^{(n)}$. By the uniqueness of the vectors $\big(\alpha_i,~1\leq i \leq \w \big)$ and $\big( \beta_e,~e\in A \big)$ in \eqref{equ3_thm_frvs_case2} for a fixed $\vv^{\,(n)}$, we further have $|\Gamma_A|=\big|\Gamma_A\big(\vv^{\,(n)}\big)\big|=q^{\w-1}$ from \eqref{frvs_5}. So, if $q>|\widetilde{\mA}_{r+1}''|$, we obtain
\begin{align*}
&\bigg|\Fq^\w \setminus \bigcup_{A\in \widetilde{\mA}_{r+1}''} \Gamma_A \bigg|=\bigg|\Fq^\w\bigg|-\bigg|\bigcup_{A\in \widetilde{\mA}_{r+1}''} \Gamma_A\bigg|
\geq  q^\w-\sum_{A\in \widetilde{\mA}_{r+1}''} |\Gamma_A|\\
&=  q^\w-q^{\w-1}\big|\widetilde{\mA}_{r+1}''\big| = q^{\w-1}\big( q-\big|\widetilde{\mA}_{r+1}''\big| \big)>0.
\end{align*}
The theorem is proved.
\end{IEEEproof}

Based on Theorem~\ref{thm_11}, for any vector $\vc=\big(c_1 \ \ c_2 \ \ \cdots \ \ c_\w \big)$ in $\Fq^\w \setminus \bigcup_{A\in \widetilde{\mA}_{r+1}''}\Gamma_A$, we let
\begin{align*}
\vb_i^{\,(n+1)}=\begin{bmatrix} \vb_i^{\,(n)}\\ c_i  \end{bmatrix},~1\leq i \leq \w, \quad \vb_i^{\,(n+1)}=\begin{bmatrix} \vb_i^{\,(n)}\\ 0  \end{bmatrix},~\w+1\leq i \leq n, \quad \text{and}\quad \vb_{n+1}^{\,(n+1)}=\begin{bmatrix} \bzero\\ 1 \end{bmatrix}.
\end{align*}
Then we let
\begin{align*}
Q^{(n+1)}\triangleq \begin{bmatrix} \vb_1^{\,(n+1)} & \vb_2^{\,(n+1)} & \cdots & \vb_{n}^{\,(n+1)} & \vb_{n+1}^{\,(n+1)} \end{bmatrix},
\end{align*}
which is an $(n+1)\times(n+1)$ matrix. Consider
\begin{align}
\Rank(Q^{(n+1)})&=\Rank\left(
\begin{bmatrix} \vb_1^{\,(n)} & \cdots & \vb_\w^{\,(n)} & \vb_{\w+1}^{\,(n)} & \cdots & \vb_n^{\,(n)} & \bzero\\ c_1 & \cdots & c_{\w} & 0 & \cdots & 0 & 1 \end{bmatrix} \right)\nonumber\\
&=\Rank\left(
\begin{bmatrix} \vb_1^{\,(n)} & \cdots & \vb_\w^{\,(n)} & \vb_{\w+1}^{\,(n)} & \cdots & \vb_n^{\,(n)} & \bzero\\ c_1 & \cdots & c_{\w} & 0 & \cdots & 0 & 1 \end{bmatrix}\cdot
\begin{bmatrix} I_n & \bzero \\ -\vc & 1 \end{bmatrix} \right)\nonumber\\
&=\Rank\left(
\begin{bmatrix} \vb_1^{\,(n)} & \cdots & \vb_\w^{\,(n)} & \vb_{\w+1}^{\,(n)} & \cdots & \vb_n^{\,(n)} & \bzero\\ 0 & \cdots & 0 & 0 & \cdots & 0 & 1 \end{bmatrix} \right)\label{equ1-b_n+1}\\
&=\Rank\left( \begin{bmatrix} Q^{(n)} & \bzero \\ \bzero{\,^\top} & 1 \end{bmatrix}\right)=n+1,\label{equ2-b_n+1}
\end{align}
where the equality~\eqref{equ1-b_n+1} follows because $\begin{bmatrix} I_n & \bzero \\ -\vc & 1 \end{bmatrix}$ is  unit lower triangular and hence invertible, and the equality \eqref{equ2-b_n+1} follows from $$\Rank(Q^{(n)})=\Rank\left(\begin{bmatrix} \vb_1^{\,(n)} & \vb_2^{\,(n)} & \cdots & \vb_n^{\,(n)}\end{bmatrix} \right)=n.$$
In other words, $Q^{(n+1)}$ is full rank. Then, $(Q^{(n+1)})^{-1}\cdot\,\mC_{n+1}$ is an $(n+1)$-dimensional SLNC which not only retains the fixed rate $\w$ and achieves the higher security level $r+1$, but also has the same local encoding kernels as the original $n$-dimensional SLNC $(Q^{(n)})^{-1} \cdot\,\mC_n$ 
at all the non-source nodes.


\subsection{An Algorithm for Code Construction}\label{sec_algo2}

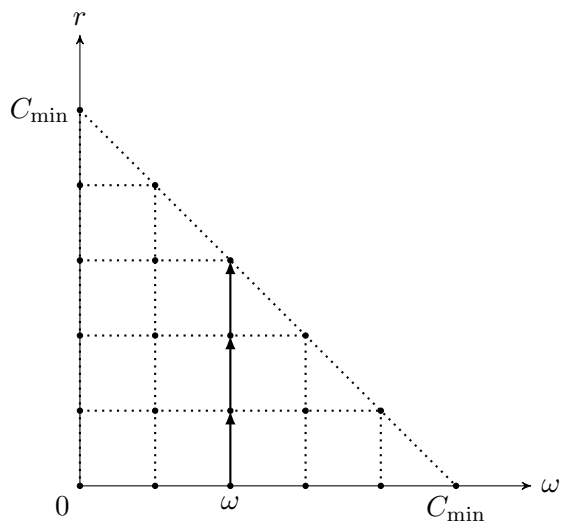
\begin{figure}[!htb]
\centering
\begin{tikzpicture}[
>=stealth',
help lines/.style={dotted, thick},
axis/.style={<->},
important line/.style={thick, -latex},
connection/.style={thick, dashed}]
    \coordinate (y) at (0,6);
    \coordinate (x) at (6,0);
    \draw[<->] (y) node[above] {$r$} -- (0,0) node [below left]{$0$} --  (x) node[right]
    {$\w$};
    \path
    coordinate (start) at (0,5)
    coordinate (p00) at (0,0)
    coordinate (p01) at (0,1)
    coordinate (p02) at (0,2)
    coordinate (p03) at (0,3)
    coordinate (p04) at (0,4)
    coordinate (p10) at (1,0)
    coordinate (p11) at (1,1)
    coordinate (p12) at (1,2)
    coordinate (p13) at (1,3)
    coordinate (p14) at (1,4)
    coordinate (p20) at (2,0)
    coordinate (p21) at (2,1)
    coordinate (p22) at (2,2)
    coordinate (p23) at (2,3)
    coordinate (p30) at (3,0)
    coordinate (p31) at (3,1)
    coordinate (p32) at (3,2)
    coordinate (p40) at (4,0)
    coordinate (p41) at (4,1)
    coordinate (end) at (5,0);
    \draw[help lines] (start) node [left] {$C_{\min}$} -- (end) node[below]
    {$C_{\min}$};
    \draw[help lines] (p10) -- (p14);

    \draw[important line] (p20)node[below]{$\w$} -- (p21);
    \draw[important line] (p21) -- (p22);
    \draw[important line] (p22) -- (p23);

    \draw[help lines] (p30) -- (p32);
    \draw[help lines] (p40) -- (p41);
    \draw[help lines] (p00) -- (start);
    \draw[help lines] (p01) -- (p41);
    \draw[help lines] (p02) -- (p32);
    \draw[help lines] (p03) -- (p23);
    \draw[help lines] (p04) -- (p14);
     \filldraw [black]
     (start) circle (1pt)
     (p00) circle (1pt)
     (p01) circle (1pt)
     (p02) circle (1pt)
     (p03) circle (1pt)
     (p04) circle (1pt)
     (p10) circle (1pt)
     (p11) circle (1pt)
     (p12) circle (1pt)
     (p13) circle (1pt)
     (p14) circle (1pt)
     (p20) circle (1pt)
     (p21) circle (1pt)
     (p22) circle (1pt)
     (p23) circle (1pt)
     (p30) circle (1pt)
     (p31) circle (1pt)
     (p32) circle (1pt)
     (p40) circle (1pt)
     (p41) circle (1pt)
     (end) circle (1pt);
\end{tikzpicture}
\caption{Local-encoding-preserving SLNCs for a fixed rate and a flexible security level.}
\label{fig-fix-Rate-flexible-SecLevel}
\end{figure}

In the last subsection, we presented an approach for designing a family of local-encoding-preserving SLNCs for a fixed information rate and multiple security levels. In particular, Theorem~\ref{thm_11} gives the prescription for designing an appropriate vector $\vc$ which is crucial for constructing a local-encoding-preserving SLNC at one security level higher. However, Theorem~\ref{thm_11} does not provide a method to find~$\vc$ readily. This is tackled in Algorithm~\ref{algo-2}, which  gives a polynomial-time implementation for constructing an $(n+1)$-dimensional SLNC with rate $\w$ and security level $r+1$ (here $n=\w+r$) from an $n$-dimensional SLNC with rate $\w$ and security level $r$, where the $(n+1)$-dimensional SLNC has the same local encoding kernels as the original $n$-dimensional SLNC at all the non-source nodes. Starting from $r=0$, by applying Algorithm~\ref{algo-2} repeatedly, we can obtain a family of local-encoding-preserving SLNCs with a fixed rate $\w$ and security levels from $0$ to $C_{\min}-\w$. This procedure is illustrated in Fig.~\ref{fig-fix-Rate-flexible-SecLevel}.

\vspace{10mm}

\begin{algorithm}[!htp]\footnotesize
\SetAlgoLined
\vspace{1mm}
{Let $\mC_{n}$ and $\mC_{n+1}$ be linear network codes defined in Lemma~\ref{thm_decoding_condition} with $q>\big|\widetilde{\mA}_{r+1}''\big|$.}\newline
\vspace{-6mm}

\KwIn{ An invertible $n \times n$ matrix $Q^{(n)}=\Big[\vb_1^{\,(n)} \ \vb_2^{\,(n)} \ \cdots \ \vb_n^{\,(n)} \Big]$ over $\Fq$ such that $(Q^{(n)})^{-1}\cdot\mC_{n}$ is an $n$-dimensional SLNC with rate $\w$ and security level $r$, where $n=\w+r<C_{\min}$.}
\KwOut{ An invertible $(n+1) \times (n+1)$ matrix $Q^{(n+1)}$ such that
$(Q^{(n+1)})^{-1}\cdot\,\mC_{n+1}$ is an $(n+1)$-dimensional SLNC with rate $\w$ and security level $r+1$.
}
\BlankLine
\Begin{
\nl Set $\mA=\emptyset$ and $\vc^{\,*}=\big(c_1^* \ \ c_2^* \ \ \cdots \ \ c_\w^* \big)=\bzero$\;
\nl\label{algo2-for}      \For{ each $A\in \mA_{r+1}$}{
\nl\label{algo2-solution} find a nonzero solution $\big(\alpha_{A,i}, 1\leq i \leq \w,\ \beta_e, e\in A \big)$ for $\sum_{i=1}^{\w}\alpha_{A,i}\vb_i^{\,(n)}=\sum_{e\in A} \beta_e\vf_e^{\,(n)}$\;
\tcp*[f]{\rm\footnotesize By Lemma~\ref{lem_verfi_algo-2}, there always exists such a nonzero solution.
}\newline
\nl\label{algo2-if_solution_nonzero} \If(\tcp*[f]{\rm\footnotesize By Lemma~\ref{lem_verfi_algo-2}, $\big(\alpha_{A,i}, 1\leq i \leq \w\big)\neq \bzero~\Leftrightarrow~A\in \widetilde{\mA}_{r+1}''$.}){$\big(\alpha_{A,i}, 1\leq i \leq \w\big)\neq \bzero$}{
\nl\label{algo2-lambda_A} compute $\lambda_A=\sum_{e\in A} \beta_e  f_{e,n+1}$\;
\nl\label{algo2-save} save the pair $\big(\alpha_{A,i}, 1\leq i \leq \w\big)$ and $\lambda_A$;
\tcp*[f]{\rm\footnotesize for the use in subsequent iterations of the ``for'' loop}\newline
\nl\label{algo2-tau_A} compute $\tau_A=\sum_{i=1}^{\w} \alpha_{A,i}\; c_i^*$\;
\nl\label{algo2-if_tau_A_zero} \If(\tcp*[f]{\rm\footnotesize Obtain a new $\vc^{\,*}$ s.t. $\sum_{i=1}^{\w} \alpha_{B,i}\; c_i^*\neq 0$, $\forall\,B\in \mA\cup \{A\}$.}){$\tau_A=0$}{
\nl\label{algo2-vh_A} find $\vh_A=\big(h_{A,1} \ \ h_{A,2} \ \ \cdots \ \ h_{A,\w} \big)\in \Fq^\w$ such that $\pi_A=\sum_{i=1}^{\w}\alpha_{A,i}\; h_{A,i}\neq 0$\;
\nl\label{algo2-if-mA_empty} \eIf{$\mA=\emptyset$}
{
\nl\label{algo2-vc*-mA_empty}  set $\vc^{\,*}=\vh_A$\;}
{
\nl\label{algo2-tau_B_pi_B} compute $\tau_B=\sum_{i=1}^{\w}\alpha_{B,i}\; c_i^*$ and  $\pi_B=\sum_{i=1}^{\w}\alpha_{B,i}\; h_{A,i}$, $\forall\,B\in\mA$\;
\tcp*[h]{\rm\footnotesize $\big(\alpha_{B,i}, 1\leq i \leq \w\big)$, $B\in\mA$ have been saved in Line~\ref{algo2-save} in previous iterations of the ``for'' loop.}\newline
\tcp*[h]{\rm\footnotesize $\tau_B=\sum_{i=1}^{\w}\alpha_{B,i}\; c_i^*\neq 0$, $\forall\,B\in\mA$ from the last iteration of the ``for'' loop.}\newline
\nl\label{algo2-xi} choose $\xi \in \Fq$ such that $\xi\cdot\tau_B + \pi_B \neq 0$, $\forall\, B\in \mA$\;
\nl\label{algo2-vc*} replace $\vc^{\,*}$ by $\xi\vc^{\,*}+\vh_A$;\tcp*[f]{\rm\footnotesize For the updated $\vc^{\,*}$, $\tau_A\neq 0$ and $\tau_B\neq 0$, $\forall\,B\in \mA$.}
}
}
\nl\label{algo2-update_mA}add $A$ into $\mA$;\tcp*[f]{\rm\footnotesize Update $\mA$. Then $\tau_B\neq 0$, $\forall\,B\in \mA$.}
}
}
\tcp*[f]{\rm\footnotesize After the ``for'' loop, $\mA=\widetilde{\mA}_{r+1}''$ and $\vc^{\,*}$ satisfies $\tau_A\neq 0 $, $\forall~A\in \mA=\widetilde{\mA}_{r+1}''$.}\newline
\nl\label{algo2-theta} choose an element $\theta$ in $\Fq$ such that $\theta\cdot\tau_A\neq \lambda_A$, $\forall~A\in \mA$\;
\nl\label{algo2-vc} calculate $\vc=\big(c_1 \ \ c_2 \ \ \cdots \ \ c_\w \big)=\theta\big(c_1^* \ \ c_2^* \ \ \cdots \ \ c_\w^* \big)=\theta\vc^{\,*}$\;
\tcp*[f]{\rm\footnotesize $\theta\cdot\tau_A \neq \lambda_A$, $\forall\,A\in\mA$ $\Rightarrow$ $\sum_{i=1}^{\w}\alpha_{A,i}\; c_i\neq \lambda_A$, $\forall\,A\in\widetilde{\mA}_{r+1}''$.}

\tcp*[f]{\rm\footnotesize This ensures that $\vc \in \Fq^\w \setminus \bigcup_{A\in \widetilde{\mA}_{r+1}''}\Gamma_A$.}\newline
\nl\label{algo2-b_i} let $\vb_i^{\,(n+1)}=\begin{bmatrix} \vb_i^{\,(n)}\\ c_i  \end{bmatrix}$, $1\leq i \leq \w$, \quad $\vb_i^{\,(n+1)}=\begin{bmatrix} \vb_i^{\,(n)}\\ 0  \end{bmatrix}$, $\w+1\leq i \leq n$, \ and \ $\vb_{n+1}^{\,(n+1)}=\begin{bmatrix} \bzero\\ 1  \end{bmatrix}$\;
\nl\label{algo2-Q-n+1} return $Q^{(n+1)}=\begin{bmatrix} \vb_1^{\,(n+1)} & \vb_2^{\,(n+1)} & \cdots & \vb_{n+1}^{\,(n+1)} \end{bmatrix}$.
}
\caption{Construction of a rate-$\w$ and security-level-$(r+1)$ SLNC from a rate-$\w$ and security-level-$r$ SLNC, both of which have the same local encoding kernels at all the non-source nodes.
}
\label{algo-2}
\end{algorithm}

\newpage

\noindent\textbf{Verification of Algorithm~\ref{algo-2}:}

For the purpose of verifying Algorithm~\ref{algo-2}, it suffices to verify that the vector $\vc=\big(c_1 \ \  c_2 \ \ \cdots \ \ c_\w \big)$ obtained in Line~\ref{algo2-vc} satisfies that $\vc \in \Fq^\w \setminus \bigcup_{A\in \widetilde{\mA}_{r+1}''}\Gamma_A$. First, we present the following lemma.

\begin{lemma}\label{lem_verfi_algo-2}
Let $A$ be any wiretap set in $\mA_{r+1}$. Then there always exists a nonzero solution
$\big(\alpha_{A,i}, 1\leq i \leq \w,\ \beta_e, e\in A \big)\in \Fq^{n+1}$ for the equation
\begin{align}\label{equ6}
\sum_{i=1}^{\w}\alpha_{A,i}\vb_i^{\,(n)}=\sum_{e\in A} \beta_e\vf_e^{\,(n)},
\end{align}
and
\begin{align}\label{equ1-lem_verfi_algo-2}
\big(\alpha_{A,i}, 1\leq i \leq \w\big)\neq \bzero~\Longleftrightarrow~A\in\widetilde{\mA}_{r+1}''.
\end{align}
\end{lemma}
\begin{IEEEproof}
The equation \eqref{equ6} can be written as the following system of linear equations
\begin{align}\label{equ1-pf-lem_verfi_algo-2}
\begin{bmatrix}\vb_i^{\,(n)},~1\leq i \leq \w,& \vf_e^{\,(n)},~e\in A \end{bmatrix}\cdot\big(\alpha_{A,i}, 1\leq i \leq \w,\ -\beta_e, e\in A \big)^\top=\bzero.
\end{align}
In view of the matrix on the LHS of the above equation of size $n\times (n+1)$, \eqref{equ1-pf-lem_verfi_algo-2} contains $n$ linear equations and $n+1$ variables (i.e., $\alpha_{A,i}$, $1\leq i \leq \w$ and $\beta_e$, $e\in A$), which implies that there must exist nonzero solutions for \eqref{equ1-pf-lem_verfi_algo-2}, or equivalently, \eqref{equ6}.

We now prove the ``only if'' part of \eqref{equ1-lem_verfi_algo-2}. Let $\big(\alpha_{A,i}, 1\leq i \leq \w,\ \beta_e, e\in A \big)$ be a nonzero solution with $\big(\alpha_{A,i}, 1\leq i \leq \w\big)\neq \bzero$ for \eqref{equ6}. Together with the linear independence of $\vb_i^{\,(n)}$, $1\leq i \leq \w$, we have
\begin{align*}
\bzero \neq \sum_{i=1}^{\w}\alpha_{A,i}\vb_i^{\,(n)}=\sum_{e\in A} \beta_e\vf_e^{\,(n)}.
\end{align*}
This immediately implies that $\mB_\w^{(n)}\cap\mL_{A}^{(n)}\neq \{\bzero\}$, proving that $A\in \widetilde{\mA}_{r+1}''$ by Lemma~\ref{lem7}.

To prove the ``if'' part of \eqref{equ1-lem_verfi_algo-2}, we assume the contrary that for a wiretap set $A\in \widetilde{\mA}_{r+1}''$, there exists a nonzero solution $\big(\alpha_{A,i}, 1\leq i \leq \w,\ \beta_e, e\in A \big)$ but $\big(\alpha_{A,i}, 1\leq i \leq \w\big)=\bzero$ for the equation \eqref{equ6}. Thus, we see that $\big(\beta_e,~e\in A \big) \neq \bzero$. Together with the linear independence of $\vf_e^{\,(n)}$, $e\in A$ (by the definition of $\widetilde{\mA}_{r+1}''$ in~\eqref{A_r+1''}), we obtain that $\sum_{e\in A} \beta_e\vf_e^{\,(n)}\neq \bzero$. On the other hand, we have $\sum_{i=1}^{\w}\alpha_{A,i}\vb_i^{\,(n)}=\bzero$ because $\big(\alpha_{A,i},~1\leq i \leq \w\big)=\bzero$. This immediately contradicts the assumption that $\big(\alpha_{A,i}, 1\leq i \leq \w,\ \beta_e, e\in A \big)$ is a solution for \eqref{equ6}. This lemma is proved.
\end{IEEEproof}

By Lemma~\ref{lem_verfi_algo-2}, it follows from Lines~\ref{algo2-solution}, \ref{algo2-if_solution_nonzero}, and \ref{algo2-update_mA} that after the ``for'' loop (Lines~\ref{algo2-for}--\ref{algo2-update_mA}), the output set $\mA$ is equal to $\widetilde{\mA}_{r+1}''$. Next, we will verify by induction that after every iteration for a wiretap set $A \in \mA_{r+1}$ (Lines~\ref{algo2-solution}--\ref{algo2-update_mA}), the condition
\begin{align}\label{equ-verif-algo2}
\tau_B=\sum_{i=1}^{\w} \alpha_{B,i}\; c_i^*\neq 0, \quad \forall~B\in \mA
\end{align}
is satisfied for the updated $\vc^{\,*}$ and the updated $\mA$. Then, upon completion of the ``for'' loop, with $\mA=\widetilde{\mA}_{r+1}''$, this implies that the vector $\vc^{\,*}$ satisfies
\begin{align}\label{equ1-verif-algo2}
\tau_A=\sum_{i=1}^{\w} \alpha_{A,i}\; c_i^*\neq 0, \quad \forall~A\in \widetilde{\mA}_{r+1}''.
\end{align}

First, we note that the condition~\eqref{equ-verif-algo2} is satisfied for $\mA=\emptyset$ and $\vc^{\,*}=\bzero$. Assume that \eqref{equ-verif-algo2} is satisfied after a number of iterations of the ``for'' loop, with at least one wiretap set $A\in\mA_{r+1}$ that has not been processed. In the next iteration, one such $A$ is processed. At this point, the set $\mA$ contains the wiretap sets $B$ in $\widetilde{\mA}_{r+1}''$ that have already been processed in the previous iterations.

\noindent\textbf{Case 1:} For the wiretap set $A$, if the nonzero solution $\big(\alpha_{A,i}, 1\leq i \leq \w,\ \beta_e, e\in A \big)$ of the equation~\eqref{equ6} found in Line~\ref{algo2-solution} satisfies $\big(\alpha_{A,i}, 1\leq i \leq \w\big)=\bzero$ (i.e., $A\in \widetilde{\mA}_{r+1}'$ by Lemma~\ref{lem_verfi_algo-2}), the ``if'' statement (Lines~\ref{algo2-if_solution_nonzero}--\ref{algo2-update_mA}) is not executed. Then, the vector $\vc^{\,*}$ and the set $\mA$ are unchanged, so that by the induction hypothesis, the condition~\eqref{equ-verif-algo2} is satisfied.

\noindent\textbf{Case 2:} Otherwise, we let $\big(\alpha_{A,i}, 1\leq i \leq \w,\ \beta_e, e\in A \big)$ with $\big(\alpha_{A,i}, 1\leq i \leq \w\big)\neq \bzero$ be the nonzero solution of the equation~\eqref{equ6} found in Line~\ref{algo2-solution} (i.e., $A\in \widetilde{\mA}_{r+1}''$ by Lemma~\ref{lem_verfi_algo-2}). For this case, the ``if'' statement (Lines~\ref{algo2-if_solution_nonzero}--\ref{algo2-update_mA}) is executed. Then, we compute $\tau_A=\sum_{i=1}^{\w} \alpha_{A,i}\; c_i^*$ and consider the following two subcases.

\textbf{Case 2A:} $\tau_A \neq 0$.

The second ``if'' statement (Lines~\ref{algo2-if_tau_A_zero}--\ref{algo2-vc*}) is not executed, so that the vector $\vc^{\,*}$ is unchanged. By Line~\ref{algo2-update_mA}, the set $\mA$ is updated by including $A$. Combining $\tau_A \neq 0$ and the induction hypothesis, \eqref{equ-verif-algo2} is satisfied for the unchanged vector $\vc^{\,*}$ and the updated set $\mA$.

\textbf{Case 2B:} $\tau_A = 0$.

The second ``if'' statement (Lines~\ref{algo2-if_tau_A_zero}--\ref{algo2-vc*}) is executed. Then, we find $\vh_A=\big(h_{A,1} \ \ h_{A,2} \ \ \cdots \ \ h_{A,\w} \big)\in \Fq^\w$ such that $\pi_A=\sum_{i=1}^{\w}\alpha_{A,i}\; h_{A,i}\neq 0$ in Line~\ref{algo2-vh_A}. If $\mA=\emptyset$ (Line~\ref{algo2-if-mA_empty}), then for this wiretap set $A$, Algorithm~\ref{algo-2} finds a nonzero solution $\big(\alpha_{A,i}, 1\leq i \leq \w,\ \beta_e, e\in A \big)$ of $\sum_{i=1}^{\w}\alpha_{A,i}\vb_i^{\,(n)}=\sum_{e\in A} \beta_e\vf_e^{\,(n)}$ with $\big(\alpha_{A,i}, 1\leq i \leq \w \big)\neq \bzero$ for the first time. On the other hand, $\vc^{\,*}=\bzero$, the initial value, so that $\tau_A=\sum_{i=1}^{\w} \alpha_{A,i}\; c_i^*=0$. Then, we update $\vc^{\,*}=\vh_A$ in Line~\ref{algo2-vc*-mA_empty} and update $\mA$ from $\emptyset$ to $\{A\}$ in Line~\ref{algo2-update_mA}. Now, we have
\begin{align*}
\tau_A=\sum_{i=1}^{\w} \alpha_{A,i}\; c_i^*=\sum_{i=1}^{\w} \alpha_{A,i}\; h_{A,i}=\pi_A\neq 0.
\end{align*}
In other words, after this iteration for $A$, the condition~\eqref{equ-verif-algo2} holds for the updated $\vc^{\,*}$ ($=\vh_A$) and the updated $\mA$~($=\{A\}$).

Otherwise, i.e., $\mA\neq \emptyset$. We compute $\tau_B=\sum_{i=1}^{\w}\alpha_{B,i}\; c_i^*$ and $\pi_B=\sum_{i=1}^{\w}\alpha_{B,i}\; h_{A,i}$ for all $B\in \mA$ in Line~\ref{algo2-tau_B_pi_B}. By Line~\ref{algo2-xi}, we choose $\xi \in \Fq$ such that $\xi\cdot\tau_B+ \pi_B \neq 0$, $\forall\, B\in \mA$, which is equivalent to choosing
$$\xi\in \Fq\setminus \bigcup_{B\in \mA}\left\{ -\frac{\pi_B}{\tau_B}\right\},$$
where we note that $\tau_B\neq 0$ for each $B\in \mA$ by the induction hypothesis. This immediately implies that the field size $q>|\widetilde{\mA}_{r+1}''|\geq |\mA|$ is sufficient for the existence of such a $\xi$. Now, we update $\vc^{\,*}$ to $\xi\vc^{\,*}+\vh_A$ in Line~\ref{algo2-vc*}. For the wiretap sets $B$ in $\mA$, it follows from Line~\ref{algo2-xi} that
\begin{align*}
\sum_{i=1}^{\w}\alpha_{B,i}\cdot(\xi c_i^*+h_{A,i})=\xi\cdot\sum_{i=1}^{\w}\alpha_{B,i}\; c_i^*+ \sum_{i=1}^{\w}\alpha_{B,i}\; h_{A,i}=\xi\cdot\tau_B+ \pi_B \neq0,
\end{align*}
and for the wiretap set $A$, it follows from $\tau_A=0$ and Line~8 that
\begin{align*}
\sum_{i=1}^{\w}\alpha_{A,i}\cdot(\xi c_i^*+h_{A,i})=\xi\cdot\sum_{i=1}^{\w}\alpha_{A,i}\; c_i^*+ \sum_{i=1}^{\w}\alpha_{A,i}\; h_{A,i}=\xi\cdot\tau_A+\pi_A=\pi_A\neq0.
\end{align*}
Thus, for this updated vector $\vc^{\,*}$ (i.e., $\xi\vc^{\,*}+\vh_A$) and the updated $\mA$ that includes $A$, \eqref{equ-verif-algo2} is satisfied.

\medskip

Therefore, we have verified that after every iteration of the ``for'' loop (Lines~\ref{algo2-solution}--\ref{algo2-update_mA}), the condition \eqref{equ-verif-algo2} is satisfied for the updated $\vc^{\,*}$ and the updated $\mA$. Finally, we verify that the vector $\vc$ obtained in Lines~\ref{algo2-theta}~and~\ref{algo2-vc} satisfies $\vc \in \Fq^{\w}\setminus \bigcup_{A\in \widetilde{\mA}_{r+1}''}\Gamma_A$. Let $A$ be an arbitrary wiretap set in $\widetilde{\mA}_{r+1}''$, and $\big(\alpha_{A,i}, 1\leq i \leq \w,\ \beta_e, e\in A \big)$ with $\big(\alpha_{A,i}, 1\leq i \leq \w\big)\neq \bzero$ be the solution found in Line~3. Let \begin{align}\label{equ2-verif-algo2}
\vv^{\,(n)}=\sum_{i=1}^{\w}\alpha_{A,i}\vb_i^{\,(n)}=\sum_{e\in A} \beta_e\vf_e^{\,(n)},
\end{align}
where we note that $\vv^{\,(n)}$ is a nonzero vector because $\big(\alpha_{A,i}, 1\leq i \leq \w\big) \neq \bzero$ and $\vb_i^{\,(n)}$, $1\leq i \leq \w$ are linearly independent. Together with the linear independence of $\vf_e^{\,(n)}$, $e\in A$ (by the definition of $\widetilde{\mA}_{r+1}''$ in~\eqref{A_r+1''}), for $\vv^{\,(n)}$ in~\eqref{equ2-verif-algo2}, both $\big(\alpha_{A,i}, 1\leq i \leq \w\big)\in\Fq^{\w}$ and $\big(\beta_e, e\in A \big)\in \Fq^{r+1}$ are unique. Consequently, by $\Gamma_A=\Gamma_A\big(\vv^{\,(n)}\big)$ (cf.~\eqref{equ14_thm_frvs_case2}) and the definition of $\Gamma_A\big(\vv^{\,(n)}\big)$ (cf.~\eqref{Gamma_v^n}), we obtain that
\begin{align*}
\vc \in \Fq^{\w} \setminus \Gamma_A~\Longleftrightarrow~\vc \in \Fq^{\w} \setminus \Gamma_A\big(\vv^{\,(n)}\big)~\Longleftrightarrow~
\sum_{i=1}^{\w}\alpha_{A,i}\; c_i \neq \sum_{e\in A} \beta_e f_{e,n+1}=\lambda_A,
\end{align*}
where the RHS is exactly the requirement in Line~\ref{algo2-theta} (with $\vc=\theta\cdot\vc^{\,*}$). Thus, we have verified that the vector $\vc$ obtained in Line~\ref{algo2-vc} satisfies that $\vc \in \Fq^\w \setminus \bigcup_{A\in \widetilde{\mA}_{r+1}''}\Gamma_A$.

It remains to verify the existence of such an element $\theta$ in Line~\ref{algo2-theta}. For each $A\in \widetilde{\mA}_{r+1}''$, we have verified that $\tau_A=\sum_{i=1}^{\w}\alpha_{A,i}\; c_i^*\neq0$ (cf.~\eqref{equ1-verif-algo2}). Thus, in order to choose $\theta \in \Fq$ satisfying $\theta\cdot\tau_A\neq \lambda_A$ for all $A\in \widetilde{\mA}_{r+1}''$, it is equivalent to choosing $\theta \in \Fq \setminus \bigcup_{A\in \widetilde{\mA}_{r+1}''}\big\{ \lambda_A\cdot\tau_A^{-1} \big\}$. This immediately implies that the field size $q>|\widetilde{\mA}_{r+1}''|$ is sufficient for the existence of such a $\theta$.

\bigskip

Next, we give an example to illustrate Algorithm~\ref{algo-2}, in which the same setup as in Example~1 in \cite{part1} is used.

\begin{eg}\label{eg-2}

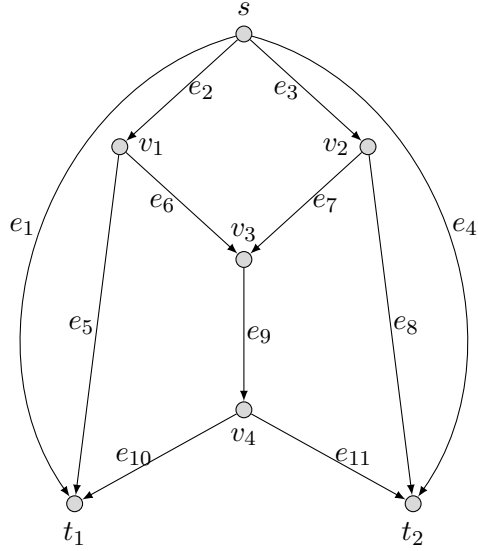
\begin{figure}[!t]
\tikzstyle{vertex}=[draw,circle,fill=gray!30,minimum size=6pt, inner sep=0pt]
  \centering
{
 \begin{tikzpicture}[x=0.6cm]
    \draw (0,0) node[vertex]     (s)[label=above:$s$] {};
    \draw (-2.75,-1.5) node[vertex] (1) [label=right:$v_1$] {};
    \draw ( 2.75,-1.5) node[vertex] (2) [label=left:$v_2$] {};
    \draw ( 0,-3) node[vertex] (3) [label=above:$v_3$] {};
    \draw ( 0,-5) node[vertex] (6) [label=below:$v_4$] {};
    \draw (-3.75,-6.25) node[vertex] (4) [label=below:$t_1$] {};
    \draw ( 3.75,-6.25) node[vertex] (5) [label=below:$t_2$] {};

    \draw[->,>=latex] (s) -- (1) node[midway, auto,swap, right=-0.5mm] {$e_2$};
    \draw[->,>=latex] (s) -- (2) node[midway, auto, left=-0.5mm] {$e_3$};
    \draw[->,>=latex] (1) -- (4) node[midway, auto,swap, left=-1mm] {$e_5$};
    \draw[->,>=latex] (1) -- (3) node[midway, auto,left=-0.5mm] {$e_6$};
    \draw[->,>=latex] (2) -- (3) node[midway, auto,right=-0.5mm] {$e_7$};
    \draw[->,>=latex] (2) -- (5) node[midway, auto, right=-1mm] {$e_8$};
    \draw[->,>=latex] (3) -- (6) node[midway, auto, right=-1mm] {$e_9$};
    \draw[->,>=latex] (6) -- (4) node[midway, auto, left=-0.5mm] {$e_{10}$};
    \draw[->,>=latex] (6) -- (5) node[midway, auto,swap, right=-0.5mm] {$e_{11}$};
    \draw[->,>=latex] (s) edge[bend right=55] node[pos=0.5,left=-0.5mm] {$e_1$} (4) ;
    \draw[->,>=latex] (s) edge[bend left=55]  node[pos=0.5,right=-0.5mm] {$e_4$} (5);
    \end{tikzpicture}
}
\caption{The network $G=(V,E)$.}
  \label{fig-part2}
\end{figure}

We consider the network $G=(V,E)$ depicted in Fig.~\ref{fig-part2}. Let
\begin{align}\label{local_kernels_EX}
\mC_3=\left\{K_{s}=\left[\begin{smallmatrix} 0 & 1 & 1 & 0 \\ 1 & 0 & 0 & 1 \\ 1 & 1 & 2 & 2 \end{smallmatrix}\right],\quad
K_{1}=K_{2}=K_{4}=\left[\begin{smallmatrix} 1 & 1 \end{smallmatrix}\right],\quad K_{3}=\big[\begin{smallmatrix} 4 \\ 1 \end{smallmatrix}\big]\right\}
\end{align}
be a $3$-dimensional linear network code over the field $\mathbb{F}_5$ on $G$, where $K_{i}$ represents the local encoding kernels at the intermediate nodes $v_i$, $1\leq i \leq 4$.
As defined in Lemma~\ref{thm_decoding_condition},
\begin{align*}
\mC_2=\begin{bmatrix}I_2 & \bzero \end{bmatrix}\cdot \mC_3 = \left\{ \begin{bmatrix}I_2 & \bzero \end{bmatrix} \cdot K_{s}=\left[\begin{smallmatrix} 0 & 1 & 1 & 0 \\ 1 & 0 & 0 & 1  \end{smallmatrix}\right],\
K_{1}=K_{2}=K_{4}=\left[\begin{smallmatrix} 1 & 1 \end{smallmatrix}\right],\ K_{3}=\big[\begin{smallmatrix} 4 \\ 1 \end{smallmatrix}\big]\right\}
\end{align*}
is a local-encoding-preserving $2$-dimensional linear network code $\mC_2$ over $\mathbb{F}_5$, of which all the global encoding kernels $\vf_i^{\,(2)}$ for $e_i$, $1\leq i \leq 11$, are
\begin{align*}
&\vf_1^{\,(2)}=\big[\begin{smallmatrix}0 \\ 1 \end{smallmatrix}\big], \quad \vf_2^{\,(2)}=\vf_5^{\,(2)}=\vf_6^{\,(2)}=\big[\begin{smallmatrix}1 \\ 0 \end{smallmatrix}\big], \quad \vf_3^{\,(2)}=\vf_7^{\,(2)}=\vf_8^{\,(2)}=\big[\begin{smallmatrix}1 \\ 0 \end{smallmatrix}\big],\\
&\vf_4^{\,(2)}=\big[\begin{smallmatrix} 0 \\ 1 \end{smallmatrix}\big], \quad \vf_9^{\,(2)}=\vf_{10}^{\,(2)}=\vf_{11}^{\,(2)}=\big[\begin{smallmatrix} 0 \\ 0 \end{smallmatrix}\big].
\end{align*}
Let $\w=1$ be the fixed rate. We first consider the security level $r=1$, and the set of the primary edge subsets of size $1$ is
\begin{align}\label{mA_1}
\mA_1=\big\{ \{e_1\}, \{e_2\}, \{e_3\}, \{e_4\}, \{e_9\}\big\}.
\end{align}
Let $Q^{(2)}=\Big[ \vb_1^{\,(2)} \ \ \vb_2^{\,(2)} \Big]
=\big[\begin{smallmatrix}  1 & 1\\ 1 & 0 \end{smallmatrix}\big]$ be a $2\times 2$ invertible matrix over $\mathbb{F}_5$. It can be verified that $(Q^{(2)})^{-1}\cdot\,\mC_2$ is a $2$-dimensional SLNC with rate $1$ and security level $1$.

In the following, we will use Algorithm~\ref{algo-2} to construct a local-encoding-preserving $3$-dimensional SLNC for the fixed rate $1$ and a higher security level~$2$. First, we give the set of the primary edge subsets of size $2$ as follows:
\begin{align}\label{mA_2}
\mA_2=\Big\{ \{e_1, e_2\}, \{e_1, e_3\},\{e_1, e_4\},\{e_1, e_9\},\{e_2, e_3\},\{e_2, e_4\},\{e_3, e_4\}, \{e_4, e_9\} \Big\}.
\end{align}
With $\vb_1^{\,(2)}=\big[\begin{smallmatrix}  1 \\ 1 \end{smallmatrix}\big]$, there exists a solution $(\alpha_{A,1},~\beta_e,~e\in A)$ of
$\alpha_{A,1} \vb_1^{\,(2)}=\sum_{e\in A}\beta_e \vf_e^{\,(2)}$ with $\alpha_{A,1}\neq 0$ (cf.~Line~\ref{algo2-solution}) if and only if $\vb_1^{\,(2)}\in \mL_A^{(2)}$. Thus, after finishing the ``for'' loop, the output $\mA$ contains the primary edge subsets $A\in \mA_2$ satisfying $\vb_1^{\,(2)}\in \mL_A^{(2)}$, i.e.,
$\mA = \big\{ \{e_1, e_2\}, \{e_1, e_3\}, \{e_2, e_4\}, \{e_3, e_4\}\big\}$ (we can verify that $\mA=\widetilde{\mA}_2''$),
and for each $A\in \mA$,
\begin{align*}
\mL_A^{(2)}=\big\langle  \vf_e^{\,(2)}:~e\in A \big\rangle=\big\langle  \big[\begin{smallmatrix}1 \\ 0 \end{smallmatrix}\big], \big[\begin{smallmatrix} 0 \\ 1 \end{smallmatrix}\big] \big\rangle.
\end{align*}

In the sequel, we use $\{i, j\}$ to represent an edge subset $\{e_i, e_j\}$ for notational simplicity. Then, we find $(\alpha_1=1,~\beta_i=\beta_j=1)$, which is a nonzero solution of
$\alpha_1 \vb_1^{\,(2)}=\beta_i \vf_i^{\,(2)}+\beta_j \vf_j^{\,(2)}$ (cf.~Line~\ref{algo2-solution}) for each $\{i, j\}\in \widetilde{\mA}_{2}''$. We compute $\lambda_{\{i,j\}}=\beta_i f_{i,3}+\beta_j f_{j,3}$ (cf.~Line~\ref{algo2-lambda_A}) for all $\{i, j\}\in \widetilde{\mA}_{2}''$ to obtain
$\lambda_{\{1,2\}}=2$, $\lambda_{\{1,3\}}=\lambda_{\{2,4\}}=3$, and $\lambda_{\{3,4\}}=4$.
According to the ``for'' loop of Algorithm~\ref{algo-2}, we can obtain $\vc^{\,*}=c_1^*=2$ that satisfies $\tau_A=\alpha_1c_1^*=2 \neq 0$, $\forall\,A\in \widetilde{\mA}_{2}''$. In Line~\ref{algo2-theta}, we choose $\theta=3$ such that $\theta\cdot\alpha_1 c_1^*=1\neq\lambda_{\{i,j\}}$, $\forall\;\{i,j\}\in \widetilde{\mA}_{2}''$, and then in Line~\ref{algo2-vc} we calculate  $\vc=c_1=\theta\cdot c_1^*=1$. In fact, we can calculate
$\Gamma_{\{1,2\}}=\{ 2 \}$, $\Gamma_{\{1,3\}}=\Gamma_{\{2,4\}}=\{ 3 \}$, and $\Gamma_{\{3,4\}}=\{ 4 \}$ by~\eqref{Gamma_A},
and thus we can verify that
\begin{align*}
\vc=1 \in \mathbb{F}_5 \setminus \bigcup_{A\in \widetilde{\mA}_{2}''} \Gamma_A=\{0,1\}.
\end{align*}
Finally, in Lines~\ref{algo2-b_i}~and~\ref{algo2-Q-n+1}, respectively, we let
$\vb_1^{\,(3)}=\left[\begin{smallmatrix} 1\\1\\1 \end{smallmatrix}\right]$, $\vb_2^{\,(3)}=\left[\begin{smallmatrix} 1\\0\\0 \end{smallmatrix}\right]$,  and $\vb_3^{\,(3)}=\left[\begin{smallmatrix} 0\\0\\1 \end{smallmatrix}\right]$ and output the $3\times 3$ invertible matrix
$Q^{(3)}=\Big[ \vb_1^{\,(3)}\ \  \vb_2^{\,(3)} \ \ \vb_3^{\,(3)} \Big]$. Then $(Q^{(3)})^{-1}\cdot\,\mC_3$ is a $3$-dimensional SLNC with rate $1$ and security level $2$ which has the same local encoding kernels as the SLNC $(Q^{(2)})^{-1}\cdot\,\mC_2$ at all the intermediate nodes.
\end{eg}

\noindent\textbf{Field Size of Algorithm~\ref{algo-2}:}

By the forgoing verification of Algorithm~\ref{algo-2}, a finite field $\Fq$ with $q>\Big|\widetilde{\mA}_{r+1}''\Big|$ is sufficient for constructing the matrix $Q^{(n+1)}$.
Therefore, for the fixed rate $\w$, the field size
\begin{align}\label{field_size_algo2}
q>\max\Big\{\Big|T\Big|,~\Big|\widetilde{\mA}_{r}''\Big|,~1\leq r \leq C_{\min}-\w \Big\}\footnotemark
\end{align}
\footnotetext{The reason for requiring $q > |T|$ here is to guarantee the existence of a $C_{\min}$-dimensional linear network code $\mC_{C_{\min}}$ on $G$.}is sufficient for constructing a family of local-encoding-preserving SLNCs with the fixed rate $\w$ and security levels from $0$ to $C_{\min}-\w$.

In addition, for a security level $r$, $1\leq r \leq C_{\min}-\w$, we note that $\max\big\{|T|, |\mA_r|\big\}$ is the best known lower bound on the required field size for the existence of an SLNC with rate $\w$ and security level $r$ (cf.~\cite{GY-SNC-Reduction}). Together with $\Big|\mA_r\Big|\geq \Big|\widetilde{\mA}_{r}''\Big|$ for each $1\leq r \leq C_{\min}-\w$, we thus see that there is no penalty at all on the field size (in terms of the best known lower bound) for constructing such a family of local-encoding-preserving SLNCs. Upon comparing with the required field size
\begin{align*}
q>\max\Big\{\big|T\big|,~\big|\mA_{r}\big|,~1\leq r \leq C_{\min}-\w \Big\}
\end{align*}
of the approach proposed at the beginning of Section~\ref{Sec_v-s-l} for constructing such a family of local-encoding-preserving SLNCs, by \eqref{field_size_algo2} we see that our approach here requires a field with a smaller size.

\bigskip

\noindent\textbf{Complexity of Algorithm~\ref{algo-2}:}

For the purpose of determining the computational complexity of Algorithm~\ref{algo-2}, we do not differentiate an addition from a multiplication over a finite field, although in general the time needed for a multiplication is much longer than that needed for an addition. We further assume that the computational complexity of each operation, i.e., an addition or a multiplication, is $\mO(1)$ regardless of the finite field.

Now, we discuss the complexity of Algorithm~\ref{algo-2}.
\begin{itemize}
    \item In Line~\ref{algo2-solution}, for each wiretap set $A\in \mA_{r+1}$, we can find a nonzero solution $\big(\alpha_{A,i}, 1\leq i \leq \w,\ \beta_e, e\in A \big)\in \Fq^{n+1}$ of the equation $\sum_{i=1}^{\w}\alpha_{A,i}\vb_i^{\,(n)}=\sum_{e\in A} \beta_e\vf_e^{\,(n)}$ by solving the system of linear equations
      \begin{align*}
      \begin{bmatrix}\vb_i^{\,(n)},~1\leq i \leq \w,& \vf_e^{\,(n)},~e\in A \end{bmatrix}\cdot\big(\alpha_{A,i}, 1\leq i \leq \w,\ -\beta_e, e\in A \big)^\top=\bzero,
      \end{align*}
       which takes at most $\mO\big(n^3\big)$ operations by Gaussian elimination.

       \item In view of the nonzero solution $\big(\alpha_{A,i}, 1\leq i \leq \w,\ \beta_e, e\in A \big)$ with $\big(\alpha_{A,i}, 1\leq i \leq \w \big)\neq \bzero$ for a wiretap set $A\in\widetilde{\mA}_{r+1}''$, we compute $\lambda_A=\sum_{e\in A}\beta_e f_{e,n+1}$ in Line~\ref{algo2-lambda_A} and $\tau_A=\sum_{i=1}^\w \alpha_{A,i}\; c_i^*$ in Line~\ref{algo2-tau_A}, which take at most $\mO\big(r+1\big)$ operations and $\mO\big(\w\big)$ operations, respectively.

       \item For Line~\ref{algo2-vh_A}, to find $\vh_A=\big(h_{A,1} \ \ h_{A,2} \ \ \cdots \ \ h_{A,\w} \big)\in \Fq^\w$ such that $\pi_A=\sum_{i=1}^{\w}\alpha_{A,i}\; h_{A,i}\neq 0$, it suffices to take $h_{A,i}=1$ for some $i$ with $\alpha_{A,i}\neq 0$, and $h_{A,j}=0$ for other $1\leq j \leq \w$ and $j\neq i$. Thus, the calculation of $\pi_B=\sum_{i=1}^{\w}\alpha_{B,i}\; h_{A,i}$ for a wiretap set $B\in\mA$ in Line~\ref{algo2-tau_B_pi_B} takes $\mO\big( 1 \big)$ operations. Further, the calculation of $\tau_B=\sum_{i=1}^\w \alpha_{B,i}\, c_i^*$ for a wiretap set $B\in\mA$ in Line~\ref{algo2-tau_B_pi_B} takes $\mO\big( \w \big)$ operations.

       \item Based on the above analyses, the complexity of Line~\ref{algo2-xi}, i.e., choosing $\xi\in \Fq$ such that $\xi\cdot\tau_B+ \pi_B \neq 0$ for all $B \in\mA$, is at most $\mO\big( \w|\mA|\big)$.

       \item For Line~\ref{algo2-vc*}, the calculation of the vector $\xi\vc^{\,*}+\vh_A$ takes at most $\mO\big( \w \big)$ operations.

       \item For the ``for'' loop (Lines~\ref{algo2-for}--\ref{algo2-update_mA}), the worst case in terms of the complexity is that $\tau_A=0$ for each execution of the ``if'' condition (Line~\ref{algo2-if_tau_A_zero}) with respect to a wiretap set $A\in \widetilde{\mA}_{r+1}''$. Combining the forgoing analyses, the total complexity of the ``for'' loop is at most $\mO\big(n^3|\mA_{r+1}|+\w\sum_{i=1}^{|\widetilde{\mA}_{r+1}''|-1}1\big)$.

       \item For Line~\ref{algo2-theta}, with the calculation of $\tau_A=\sum_{i=1}^\w \alpha_{A,i}\, c_i^*$ taking $\mO\big( \w \big)$ operations, the complexity of choosing $\theta \in \Fq$ such that $\theta\cdot \tau_A \neq \lambda_A$ for all $A\in \widetilde{\mA}_{r+1}''$ is at most $\mO\big( \w|\widetilde{\mA}_{r+1}''|\big)$.

       \item For Line~\ref{algo2-vc}, the calculation of $\vc=\theta\vc^{\,*}$ takes at most $\mO\big( \w \big)$ operations.
\end{itemize}

Therefore, by combining all the foregoing analyses, the total complexity of Algorithm~\ref{algo-2} is not larger than
$$\mO\Big(n^3|\mA_{r+1}|+\w|\widetilde{\mA}_{r+1}''|^2\Big).$$
With this complexity, we see that the complexity of Algorithm~\ref{algo-2} even for the worst case is considerably smaller than one $\w$th of the complexity
$$\mO\left( \w n^3\big|\mA_{r+1}\big|+\w n\big|\mA_{r+1}\big|^2+(r+1)n^3 \right)$$ (cf.~the~\ref{footnote}th footnote or Appendix~A in \cite{part1})
of the approach proposed at the beginning of Section~\ref{Sec_v-s-l}.

On the other hand, for our approach here, in order to store the matrix $Q^{(n+1)}$ at the source node $s$, it suffices to store the row $\w$-vector $\vc$ only. This implies that the storage cost is $\mO\big( \w \big)$, which is independent of the dimensions of linear network codes and considerably smaller than the storage cost $\mO\big( (n+1)^2 \big)$ of the approach proposed at the beginning of Section~\ref{Sec_v-s-l}. Thus, our approach here reduces considerably the complexity and storage cost further.

\section{Fixed-Dimension Secure Network Coding}\label{sec_fix_dim}

\begin{figure}[!t]
\centering
\begin{tikzpicture}[
>=stealth',
help lines/.style={dotted, thick},
axis/.style={<->},
important line/.style={thick},
connection/.style={thick, dashed}]
    \coordinate (y) at (0,6);
    \coordinate (x) at (6,0);
    \draw[<->] (y) node[above] {$r$} -- (0,0) node [below left]{$0$} --  (x) node[right]
    {$\w$};
    \path
    coordinate (start) at (0,5)
    coordinate (p00) at (0,0)
    coordinate (p01) at (0,1)
    coordinate (p02) at (0,2)
    coordinate (p03) at (0,3)
    coordinate (p04) at (0,4)
    coordinate (p10) at (1,0)
    coordinate (p11) at (1,1)
    coordinate (p12) at (1,2)
    coordinate (p13) at (1,3)
    coordinate (p14) at (1,4)
    coordinate (p20) at (2,0)
    coordinate (p21) at (2,1)
    coordinate (p22) at (2,2)
    coordinate (p23) at (2,3)
    coordinate (p30) at (3,0)
    coordinate (p31) at (3,1)
    coordinate (p32) at (3,2)
    coordinate (p40) at (4,0)
    coordinate (p41) at (4,1)
    coordinate (end) at (5,0);
    \draw[help lines] (start) node [left] {$C_{\min}$} -- (end) node[below] {$C_{\min}$};
    \draw[important line] (p03) node [left] {$n$} -- (p30) node[below] {$n$};
    \draw[help lines] (p10) -- (p14);
    \draw[help lines] (p20) -- (p23);
    \draw[help lines] (p30) -- (p32);
    \draw[help lines] (p40) -- (p41);
    \draw[help lines] (p01) -- (p41);
    \draw[help lines] (p02) -- (p32);
    \draw[help lines] (p03) -- (p23);
    \draw[help lines] (p04) -- (p14);
     \filldraw [black]
     (start) circle (1pt)
     (p00) circle (1pt)
     (p01) circle (1pt)
     (p02) circle (1pt)
     (p03) circle (1pt)
     (p04) circle (1pt)
     (p10) circle (1pt)
     (p11) circle (1pt)
     (p12) circle (1pt)
     (p13) circle (1pt)
     (p14) circle (1pt)
     (p20) circle (1pt)
     (p21) circle (1pt)
     (p22) circle (1pt)
     (p23) circle (1pt)
     (p30) circle (1pt)
     (p31) circle (1pt)
     (p32) circle (1pt)
     (p40) circle (1pt)
     (p41) circle (1pt)
     (end) circle (1pt);
\end{tikzpicture}
\caption{SLNCs with a fixed dimension $n$, $1\leq n \leq C_{\min}$.}
\label{fig-Line3}
\end{figure}
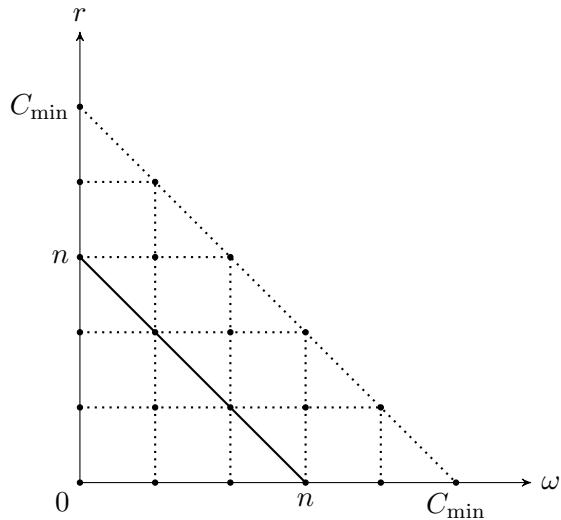

In this section, we consider the problem of designing a family of local-encoding-preserving SLNCs with a fixed dimension $n$ ($0 \leq n\leq C_{\min}$), i.e., a family of $n$-dimensional local-encoding-preserving SLNCs with the rate and security-level pair $(\w,r)$ satisfying $\w+r=n$. These pairs are all the nonnegative integer points on the line $\w+r=n$, as shown in Fig.~\ref{fig-Line3}. Note that the pair $(0,r)$ is always achievable for $0\leq r \leq C_{\min}$.

Based on the SLNC construction at the end of Section~\ref{Sec_Preli}, we naturally put forward the following approach to solve this problem. First, we construct an $n$-dimensional linear network code $\mC_n$. With this code $\mC_n$, for each nontrivial security level $1 \leq r \leq n-1$, we design an $n\times n$ invertible matrix $Q_r^{(n)}$ such that the condition~\eqref{secure_condition} is satisfied (here we use $Q_r^{(n)}$ in place of the matrix $Q^{(n)}$ in the SLNC construction at the end of Section~\ref{Sec_Preli}). Then, we obtain a family of $n$-dimensional local-encoding-preserving SLNCs $\Big\{ \big(Q_r^{(n)}\big)^{-1}\cdot\,\mC_n:~0\leq r \leq n-1 \Big\}$ (here $Q_0^{(n)}$ can be taken as the $n\times n$ identity matrix) with rate and security-level pairs $(n-r,r)$ for $0 \leq r \leq n-1$. However, the above approach not only requires the construction of the matrix $Q_r^{(n)}$ for each $r$ but also requires the source node $s$ to store all the matrices $Q_r^{(n)}$. To avoid these shortcomings, we present the following more efficient approach to construct an $n$-dimensional SLNC such that with the same SLNC, all the pairs $(\w, r)$ on the line $\w+r=n$ can be achieved. The next theorem asserts the existence of such an SLNC.

\begin{thm}\label{thm_v-r-v-sl}
Let $n$ be a nonnegative integer with $n \leq C_{\min}$, and $\mC_n$ be an $n$-dimensional linear network code on the network $G$ over a finite field $\Fq$ with \begin{align*}
q>\max\big\{|T|,~|\mA_r|,~1\leq r \leq n-1\big\},
\end{align*}
of which all the global encoding kernels are $\vf^{\,(n)}_e$, $e\in E$. Then, there exists an $n\times n$ invertible matrix $Q^{(n)}=\Big[ \vb^{\,(n)}_1 \ \ \vb^{\,(n)}_2 \ \ \cdots \ \ \vb^{\,(n)}_n \Big]$ over $\Fq$ such that for each $0\leq r \leq n$, the following condition is satisfied:
\begin{align}\label{equ_thm_multi_rate_&_s-l}
\big\langle \vb^{\,(n)}_1,\ \vb^{\,(n)}_2,\ \cdots,\ \vb^{\,(n)}_{n-r} \big\rangle \bigcap \big\langle  \vf^{\,(n)}_e:~e\in A \big\rangle=\{\bzero\}, \quad \forall~A\in \mA_{r}.\footnotemark
\end{align}
\footnotetext{For notational convenience, we let $\mA_{0}=\emptyset$ and $\big\{ \vb^{\,(n)}_1, \vb^{\,(n)}_2,\cdots, \vb^{\,(n)}_{i} \big\}=\emptyset$ for $i=0$. Then, the condition~\eqref{equ_thm_multi_rate_&_s-l} is always satisfied for $r=0$ and $r=n$.}In other words, $(Q^{(n)})^{-1}\cdot\,\mC_n$ is an $\Fq$-valued $n$-dimensional SLNC applicable to any rate and security-level pair $(n-r,r)$ for $r=0,1,2,\cdots,n$.
\end{thm}

\begin{IEEEproof}
If there exists an $n\times n$ invertible matrix $Q^{(n)}=\Big[ \vb^{\,(n)}_1 \ \ \vb^{\,(n)}_2 \ \ \cdots \ \ \vb^{\,(n)}_n \Big]$ over $\Fq$ satisfying the condition~\eqref{equ_thm_multi_rate_&_s-l} for each $0 \leq r \leq n$, then by Lemma~\ref{lemma_iff_cond}, $(Q^{(n)})^{-1}\cdot\,\mC_n$ is an $\Fq$-valued $n$-dimensional SLNC applicable to any rate and security-level pair $(n-r,r)$ for $0\leq r \leq n$. Thus, it remains to prove that if $q>\max\big\{|T|,~|\mA_r|,~1\leq r \leq n-1\big\}$, there exists such an $n\times n$ invertible matrix $Q^{(n)}$ over $\Fq$. Toward this end, we choose the $n$ vectors $\vb^{\,(n)}_1$, $\vb^{\,(n)}_2$, $\cdots$, $\vb^{\,(n)}_n$ sequentially such that the following is satisfied:
\begin{align}\label{vb_i}
\vb^{\,(n)}_i \in
\Fq^n \setminus \bigcup_{A \in \mA_{n-i}} \big(\mB_{i-1}^{(n)}+\mL_A^{(n)}\big), \qquad 1 \leq i \leq n.
\end{align}
Then we prove that $Q^{(n)}=\Big[ \vb^{\,(n)}_1 \ \ \vb^{\,(n)}_2 \ \ \cdots \ \ \vb^{\,(n)}_n \Big]$ satisfies the condition~\eqref{equ_thm_multi_rate_&_s-l}.

We first prove that all the sets on the RHS of~\eqref{vb_i} are nonempty provided that $q>\max\big\{|T|,~|\mA_r|,~1\leq r \leq n-1\big\}$. For any $1\leq i \leq n$, we consider
\begin{align*}
&\bigg|\Fq^n \setminus \bigcup_{A \in \mA_{n-i}} \big(\mB_{i-1}^{(n)}+\mL_A^{(n)}\big) \bigg|\nonumber\\
&=\bigg|\Fq^n \bigg|-\bigg| \bigcup_{A \in \mA_{n-i}} \big(\mB_{i-1}^{(n)}+\mL_A^{(n)}\big) \bigg|\nonumber\\
&\geq q^n-\sum_{A\in \mA_{n-i}} \Big|\mB_{i-1}^{(n)}+\mL_A^{(n)} \Big|\nonumber\\
& \geq q^n-\sum_{A\in \mA_{n-i}} q^{n-1}\\
& =q^{n-1}(q-|\mA_{n-i}|)>0,\nonumber
\end{align*}
where the last inequality follows because for any $1\leq i \leq n$ and $A\in\mA_{n-i}$,
\begin{align*}
\dim\big(\mB_{i-1}^{(n)}+\mL_A^{(n)}\big)\leq \dim\big(\mB_{i-1}^{(n)}\big)+\dim\big(\mL_A^{(n)}\big)=i-1+n-i=n-1.
\end{align*}

Now, by~\eqref{vb_i}, we see that
\begin{align*}
\vb^{\,(n)}_1\neq \bzero \quad \text{ and }\quad  \vb^{\,(n)}_i \in \Fq^n \setminus \mB_{i-1}^{(n)},~~2\leq i \leq n,
\end{align*}
which immediately implies that $\vb^{\,(n)}_i$, $1\leq i \leq n$ are linearly independent. Next, we will prove that for each $0\leq i \leq n$, the condition
\begin{align*}
\big\langle \vb^{\,(n)}_1,\ \vb^{\,(n)}_2,\ \cdots,\ \vb^{\,(n)}_{i} \big\rangle \bigcap \big\langle  \vf^{\,(n)}_e:~e\in A \big\rangle=\{\bzero\}, \qquad \forall~A\in \mA_{n-i},
\end{align*}
i.e.,
\begin{align}\label{equ1-thm_v-r-v-sl}
\mB_{i}^{(n)}\bigcap\mL_A^{(n)}=\{\bzero\}, \qquad \forall~A\in \mA_{n-i}
\end{align}
is satisfied, which is equivalent to~\eqref{equ_thm_multi_rate_&_s-l} by taking $r=n-i$ for $0 \leq i \leq n$. The condition~\eqref{equ1-thm_v-r-v-sl} is clearly satisfied for $i=0$ and $i=n$. It remains to prove that~\eqref{equ1-thm_v-r-v-sl} is satisfied for $1\leq i \leq n-1$, which will be done by induction as follows.

We now assume that~\eqref{equ1-thm_v-r-v-sl} is satisfied for $i-1$ and will prove that \eqref{equ1-thm_v-r-v-sl} is satisfied for $i$, $1 \leq i \leq n-1$. Suppose the contrary that \eqref{equ1-thm_v-r-v-sl} is not satisfied for $i$, namely that there exists a wiretap set $A$ in $\mA_{n-i}$ such that $\mB_{i}^{(n)} \bigcap \mL_A^{(n)} \neq \{\bzero\}$. Then there exist two nonzero vectors
$\big(\alpha_j,~1 \leq j \leq i \big) \in \Fq^i$ and $\big( \beta_e,~e\in A \big) \in \Fq^{n-i}$ such that
\begin{align}\label{equ4-thm_v-r-v-sl}
\bzero \neq \sum_{j=1}^{i}\alpha_j\vb_j^{\,(n)}=\sum_{e\in A}\beta_e \vf_e^{\,(n)},
\end{align}
where in the above $\sum_{j=1}^{i}\alpha_j\vb_j^{\,(n)} \neq \bzero$ because $\vb^{\,(n)}_j$, $1\leq j \leq i$ are linearly independent. We first claim that $\alpha_i=0$ because otherwise
\begin{align*}
\vb_i^{\,(n)}=(\alpha_i)^{-1}\cdot\bigg(\sum_{e\in A}\beta_e \vf_e^{\,(n)}-
\sum_{j=1}^{i-1}\alpha_j\vb_j^{\,(n)}\bigg) \in \mB_{i-1}^{(n)}+\mL_A^{(n)},
\end{align*}
contradicting the condition~\eqref{vb_i} that must be satisfied by $\vb_i^{\,(n)}$. For $i=1$, we have $\alpha_1=0$ which contradicts that $\big( \alpha_1 \big)$ is a nonzero vector. This proves~\eqref{equ1-thm_v-r-v-sl} for $i=1$. For $i\geq 2$, we can rewrite \eqref{equ4-thm_v-r-v-sl} as
\begin{align*}
\bzero \neq \sum_{j=1}^{i-1}\alpha_j\vb_j^{\,(n)}=\sum_{e\in A}\beta_e \vf_e^{\,(n)},
\end{align*}
implying that
\begin{align}\label{equ5-thm_v-r-v-sl}
\mB_{i-1}^{(n)} \bigcap \mL_A^{(n)}\neq \{\bzero\}.
\end{align}
Let $e\in E$ be an edge not in $A$. Such an edge $e$ exists since $|E|\geq C_{\min} \geq n > n-i = |A|$ for any $1 \leq i \leq n-1$. Then we have
$$A\cup\{e\} \in \mE_{n-(i-1)}=\big\{ A\subseteq E:\ |A|\leq n-(i-1) \big\}$$
and $\mB_{i-1}^{(n)} \bigcap \mL_{A\bigcup\{e\}}^{(n)}\neq \{\bzero\}$ by~\eqref{equ5-thm_v-r-v-sl}. By Lemma~\ref{lem_equi_Er_vs_Ar}, there must exist a wiretap set $A'\in \mA_{n-(i-1)}$ such that $\mB_{i-1}^{(n)} \bigcap \mL_{A'}^{(n)}\neq \{\bzero\}$, which is a contradiction to the induction hypothesis that~\eqref{equ1-thm_v-r-v-sl} is satisfied for $i-1$. This proves that~\eqref{equ1-thm_v-r-v-sl} is satisfied for $i$, and the theorem follows.
\end{IEEEproof}
\begin{algorithm}[!h]
\SetAlgoLined
\vspace{2mm}
\KwIn{ An $n$-dimensional linear network code $\mC_n$ over a finite field $\Fq$ with $q>\max\big\{|T|,~|\mA_r|,~1\leq r \leq n-1\big\}$.}
\KwOut{ An $n\times n$ invertible matrix $Q^{(n)}$ such that $(Q^{(n)})^{-1}\cdot\,\mC_{n}$ is an $n$-dimensional SLNC applicable to all rate and security-level pairs $(\w,r)$ with $\w+r=n$.}
\BlankLine
\Begin{
\nl choose $\vb^{\,(n)}_1 \in \Fq^n \setminus \bigcup_{A\in \mA_{n-1}} \mL_A^{(n)}$\;
\For{$i=2$ \KwTo $n-1$}{
\nl choose $\vb^{\,(n)}_i \in \Fq^n \setminus \bigcup_{A \in \mA_{n-i}} \big(\mB_{i-1}^{(n)}+\mL_A^{(n)}\big)$\;
}
\nl choose $\vb^{\,(n)}_n \in \Fq^n \setminus \mB_{n-1}^{(n)}$\;
\nl return $Q^{(n)}=\Big[ \vb^{\,(n)}_1 \ \ \vb^{\,(n)}_2 \ \ \cdots \ \ \vb^{\,(n)}_n \Big]$.
}
\caption{Construction of an $n$-dimensional SLNC applicable to all rate and security-level pairs $(\w,r)$ with $\w+r=n$.}
\label{algo-3}
\end{algorithm}

Algorithm~\ref{algo-3} is an implementation for the construction of the SLNC in the proof of Theorem~\ref{thm_v-r-v-sl}. We note that $\max\big\{|T|, |\mA_r|\big\}$ is the best known lower bound on the required field size for the existence of an SLNC with security level $r$ (cf.~\cite{GY-SNC-Reduction}). Thus, for Algorithm~\ref{algo-3}, there is no penalty on the field size (in terms of the best known lower bound) for constructing such an SLNC applicable to all rate and security-level pairs $(\w,r)$ with $\w+r=n$.

Next, we discuss the complexity of Algorithm~\ref{algo-3}. Similar to the previous complexity analyses, we do not differentiate an addition from a multiplication over a finite field, and instead we assume that the computational complexity of each operation is $\mO(1)$ regardless of the finite field. In Lines~1~and~2, the vector $\vb^{\,(n)}_i$ for $1 \leq i \leq n-1$ can be found in time $\mO\big( n^3\big|\mA_{n-i}\big|+n\big|\mA_{n-i}\big|^2 \big)$, and the vector $\vb^{\,(n)}_n$ in Line~3 can be found in at most $\mO\big( n^3+n \big)$ operations (cf.~the complexity analysis of Line~3 in Algorithm~1 in \cite{part1} or Appendix~A in \cite{part1}). Combining the above analyses, the total complexity of Algorithm~\ref{algo-3} is thus at most
\begin{align}\label{complexity-algo-3}
\mO\left( n^3\Big(1+ \sum_{i=1}^{n-1} \big|\mA_{n-i}\big|\Big)+n\Big(1+ \sum_{i=1}^{n-1} \big|\mA_{n-i}\big|^2\Big)\right).
\end{align}

For the approach proposed at the beginning of this section, we note that the computational complexity of the construction of $Q_r^{(n)}$ is
$\mO\big( (n-r)n^3|\mA_{r}|+(n-r) n |\mA_{r}|^2+r n^2 \big)$ for $1\leq r \leq n-1$  (cf.~Appendix~A in~\cite{part1}), and thus the total complexity of the construction of all the $n-1$ matrices $Q_r^{(n)}$, $1\leq r \leq n-1$ is
\begin{align*}
\mO\left( n^3\Big(n+ \sum_{r=1}^{n-1} (n-r)\big|\mA_{r}\big|\Big)+n\Big(n+ \sum_{r=1}^{n-1} (n-r)\big|\mA_{r}\big|^2 \Big) \right),
\end{align*}
or equivalently,
\begin{align}\label{complexity-pre-algo-3}
\mO\left( n^3\Big(n+ \sum_{i=1}^{n-1}\, i\,\big|\mA_{n-i}\big|\Big)+n\Big(n+ \sum_{i=1}^{n-1}\, i\,\big|\mA_{n-i}\big|^2\Big) \right).
\end{align}
Comparing~\eqref{complexity-algo-3} and~\eqref{complexity-pre-algo-3}, we see that the complexity of Algorithm~\ref{algo-3} is considerably smaller than that of the approach proposed at the beginning of this section. On the other hand, for our approach here, it suffices to store the matrix $Q^{(n)}$ only at the source node $s$, which implies that the storage cost is at most~$\mO\big( n^2 \big)$. This is also considerably smaller than the storage cost $\mO\big( n^3 \big)$ of the approach proposed at the beginning of this section that needs to store all the matrices $Q_r^{(n)}$, $1\leq i \leq n-1$.

\bigskip

We continue to use the setup in Example~\ref{eg-2} to illustrate Algorithm~\ref{algo-3}.

\begin{eg}
We consider the network $G$ (cf.~Fig.~\ref{fig-part2}) and the $3$-dimensional $\mathbb{F}_5$-valued linear network code $\mC_3$ (cf.~\eqref{local_kernels_EX}) in Example~\ref{eg-2}. Next, we use Algorithm~\ref{algo-3} to construct a $3$-dimensional SLNC $(Q^{(3)})^{-1}\cdot\,\mC_3$ achieving all $4$ rate and security-level pairs $(3,0)$, $(2,1)$, $(1,2)$ and $(0,3)$ on the line $\w+r=3$, where we remark that the pair $(0,3)$ is always achievable. For the nontrivial security levels $1$ and $2$, with the linear network code~$\mC_3$, we calculate that
\begin{itemize}
  \item according to the set $\mA_1$ of the primary edge subsets with size $1$ (cf.~\eqref{mA_1}),
\begin{align*}
\mL_{\{e_1\}}^{(3)}=\left\langle \left[\begin{smallmatrix} 0\\1\\1\end{smallmatrix}\right]\right\rangle,\
\mL_{\{e_2\}}^{(3)}=\left\langle \left[\begin{smallmatrix} 1\\0\\1\end{smallmatrix}\right]\right\rangle,\
\mL_{\{e_3\}}^{(3)}=\left\langle \left[\begin{smallmatrix} 1\\0\\2\end{smallmatrix}\right]\right\rangle,\
\mL_{\{e_4\}}^{(3)}=\left\langle \left[\begin{smallmatrix}
0\\1\\2\end{smallmatrix}\right]\right\rangle,\
\mL_{\{e_9\}}^{(3)}=\left\langle \left[\begin{smallmatrix} 0\\0\\1\end{smallmatrix}\right]\right\rangle;
\end{align*}
  \item according to the set $\mA_2$ of the primary edge subsets with size $2$ (cf.~\eqref{mA_2}),
  \begin{align*}
&\mL_{\{e_1,e_2\}}^{(3)}=\left\langle \left[\begin{smallmatrix} 0\\1\\1\end{smallmatrix}\right], \left[\begin{smallmatrix} 1\\0\\1\end{smallmatrix}\right] \right\rangle,\
\mL_{\{e_1,e_3\}}^{(3)}=\left\langle \left[\begin{smallmatrix} 0\\1\\1\end{smallmatrix}\right], \left[\begin{smallmatrix} 1\\0\\2\end{smallmatrix}\right] \right\rangle,\ \mL_{\{e_1,e_4\}}^{(3)}=\mL_{\{e_1,e_9\}}^{(3)}=\mL_{\{e_4,e_9\}}^{(3)}=\left\langle \left[\begin{smallmatrix} 0\\1\\1\end{smallmatrix}\right], \left[\begin{smallmatrix} 0\\0\\1\end{smallmatrix}\right] \right\rangle,\\
&\mL_{\{e_2,e_3\}}^{(3)}=\left\langle \left[\begin{smallmatrix} 1\\0\\1\end{smallmatrix}\right], \left[\begin{smallmatrix} 0\\0\\1\end{smallmatrix}\right] \right\rangle,\
\mL_{\{e_2,e_4\}}^{(3)}=\left\langle \left[\begin{smallmatrix} 1\\0\\1\end{smallmatrix}\right], \left[\begin{smallmatrix} 0\\1\\2\end{smallmatrix}\right] \right\rangle,\
\mL_{\{e_3,e_4\}}^{(3)}=\left\langle \left[\begin{smallmatrix} 1\\0\\2\end{smallmatrix}\right], \left[\begin{smallmatrix} 0\\1\\2\end{smallmatrix}\right] \right\rangle.
\end{align*}
\end{itemize}
In Lines~1--3, we sequentially choose column $3$-vectors $\vb^{\,(3)}_1$, $\vb^{\,(3)}_2$, $\vb^{\,(3)}_3$ as follows:
\begin{align*}
&\vb^{\,(3)}_1=\left[\begin{smallmatrix} 1\\1\\0 \end{smallmatrix}\right] \in \mathbb{F}_5^3 \setminus \bigcup_{A\in \mA_{2}} \mL^{(3)}_A,\\
&\vb^{\,(3)}_2=\left[\begin{smallmatrix} 0\\1\\0 \end{smallmatrix}\right] \in \mathbb{F}_5^3 \setminus \bigcup_{A\in \mA_{1}} \big(\mB_{1}^{(3)}+\mL_A^{(3)}\big)
=\mathbb{F}_5^3 \setminus \bigcup_{A\in \mA_{1}} \big\langle \vb^{\,(3)}_1,~\vf^{\,(3)}_e,~e\in A \big\rangle,\\
&\vb^{\,(3)}_3=\left[\begin{smallmatrix} 0\\0\\1 \end{smallmatrix}\right] \in \mathbb{F}_5^3 \setminus \mB_{2}^{(3)} = \mathbb{F}_5^3 \setminus \big\langle \vb^{\,(3)}_1,~\vb^{\,(3)}_2 \big\rangle,
\end{align*}
and output the invertible $Q^{(3)}=\Big[ \vb^{\,(3)}_1 \ \ \vb^{\,(3)}_2 \ \ \vb^{\,(3)}_3 \Big] = \left[\begin{smallmatrix} 1 & 0 & 0 \\ 1 & 1 & 0\\ 0 & 0 & 1\end{smallmatrix}\right]$. Now, $(Q^{(3)})^{-1}\cdot\,\mC_3$ is an $\mathbb{F}_5$-valued $3$-dimensional SLNC that is applicable to the rate and security-level pairs $(2,1)$ and $(1,2)$. Clearly, $(Q^{(3)})^{-1}\cdot\,\mC_3$ is also applicable to the rate and security-level pair $(3,0)$ by the invertibility of $Q^{(3)}$ and Proposition~\ref{prop}.

Theorem~\ref{thm_v-r-v-sl} guarantees the existence of an SLNC for the current example if
\begin{align*}
|\Fq|>\max\big\{ |T|,~|\mA_1|,~|\mA_2| \big\}=8.
\end{align*}
However, such an SLNC may exist in a base field with size less than or equal to $8$, which is the case for the SLNC constructed here.

\end{eg}


\section{Secure Network Coding for Flexible Rate and Security Level} \label{Sec_ultimate}

\begin{figure}[!b]
\centering
\begin{minipage}[b]{0.5\textwidth}
\centering
\begin{tikzpicture}[
>=stealth',
help lines/.style={dotted, thick},
axis/.style={<->},
important line/.style={thick, latex-},
important line 2/.style={thick},
connection/.style={thick, dashed}]
    \coordinate (y) at (0,6);
    \coordinate (x) at (6,0);
    \draw[<->] (y) node[above] {$r$} -- (0,0) node [below left]{$0$} --  (x) node[right]
    {$\w$};
    \path
    coordinate (start) at (0,5)
    coordinate (p00) at (0,0)
    coordinate (p01) at (0,1)
    coordinate (p02) at (0,2)
    coordinate (p03) at (0,3)
    coordinate (p04) at (0,4)
    coordinate (p10) at (1,0)
    coordinate (p11) at (1,1)
    coordinate (p12) at (1,2)
    coordinate (p13) at (1,3)
    coordinate (p14) at (1,4)
    coordinate (p20) at (2,0)
    coordinate (p21) at (2,1)
    coordinate (p22) at (2,2)
    coordinate (p23) at (2,3)
    coordinate (p30) at (3,0)
    coordinate (p31) at (3,1)
    coordinate (p32) at (3,2)
    coordinate (p40) at (4,0)
    coordinate (p41) at (4,1)
    coordinate (end) at (5,0);
    \draw[help lines] (start) node [left] {$C_{\min}$} -- (end) node[below] {$C_{\min}$};
    \draw[help lines] (p10) -- (p14);
    \draw[help lines] (p20) -- (p23);
    \draw[help lines] (p30) -- (p32);
    \draw[help lines] (p40) -- (p41);
    \draw[help lines] (p01) -- (p41);
    \draw[help lines] (p02) -- (p32);
    \draw[help lines] (p03) -- (p23);
    \draw[help lines] (p04) -- (p14);
     \filldraw [black]
     (start) circle (1pt)
     (p00) circle (1pt)
     (p01) circle (1pt)
     (p02) circle (1pt)
     (p03) circle (1pt)
     (p04) circle (1pt)
     (p10) circle (1pt)
     (p11) circle (1pt)
     (p12) circle (1pt)
     (p13) circle (1pt)
     (p14) circle (1pt)
     (p20) circle (1pt)
     (p21) circle (1pt)
     (p22) circle (1pt)
     (p23) circle (1pt)
     (p30) circle (1pt)
     (p31) circle (1pt)
     (p32) circle (1pt)
     (p40) circle (1pt)
     (p41) circle (1pt)
     (end) circle (1pt);
\end{tikzpicture}
\caption{The rate and security-level region.}
\label{fig-region}
\end{minipage}%
\centering
\begin{minipage}[b]{0.5\textwidth}
\centering
\begin{tikzpicture}[
>=stealth',
help lines/.style={dotted, thick},
axis/.style={<->},
important line/.style={thick, latex-},
important line 2/.style={thick},
connection/.style={thick, dashed}]
    \coordinate (y) at (0,6);
    \coordinate (x) at (6,0);
    \draw[<->] (y) node[above] {$r$} -- (0,0) node [below left]{$0$} --  (x) node[right]
    {$\w$};
    \path
    coordinate (p05) at (0,5)
    coordinate (p00) at (0,0)
    coordinate (p01) at (0,1)
    coordinate (p02) at (0,2)
    coordinate (p03) at (0,3)
    coordinate (p04) at (0,4)
    coordinate (p10) at (1,0)
    coordinate (p11) at (1,1)
    coordinate (p12) at (1,2)
    coordinate (p13) at (1,3)
    coordinate (p14) at (1,4)
    coordinate (p20) at (2,0)
    coordinate (p21) at (2,1)
    coordinate (p22) at (2,2)
    coordinate (p23) at (2,3)
    coordinate (p30) at (3,0)
    coordinate (p31) at (3,1)
    coordinate (p32) at (3,2)
    coordinate (p40) at (4,0)
    coordinate (p41) at (4,1)
    coordinate (p50) at (5,0);

    \draw[important line 2] (p05) node [left] {$C_{\min}$} --  (p50) node[below] {$C_{\min}$};

    \draw[important line] (p40) node [left] {} -- (p50) node[below] {};
    \draw[important line] (p30) node [left] {} -- (p40) node[below] {};
    \draw[important line] (p20) node [left] {} -- (p30) node[below] {};
    \draw[important line] (p10) node [left] {} -- (p20) node[below] {};
    \draw[important line] (p00) node [left] {} -- (p10) node[below] {};

    \draw[important line] (p31) node [left] {} -- (p41) node[below] {};
    \draw[important line] (p21) node [left] {} -- (p31) node[below] {};
    \draw[important line] (p11) node [left] {} -- (p21) node[below] {};
    \draw[important line] (p01) node [left] {} -- (p11) node[below] {};

    \draw[important line] (p22) node [left] {} -- (p32) node[below] {};
    \draw[important line] (p12) node [left] {} -- (p22) node[below] {};
    \draw[important line] (p02) node [left] {} -- (p12) node[below] {};

    \draw[important line] (p13) node [left] {} -- (p23) node[below] {};
    \draw[important line] (p03) node [left] {} -- (p13) node[below] {};

    \draw[important line] (p04) node [left] {} -- (p14) node[below] {};

    \draw[help lines] (p10) -- (p14);
    \draw[help lines] (p20) -- (p23);
    \draw[help lines] (p30) -- (p32);
    \draw[help lines] (p40) -- (p41);
     \filldraw [black]
     (p05) circle (1pt)
     (p00) circle (1pt)
     (p01) circle (1pt)
     (p02) circle (1pt)
     (p03) circle (1pt)
     (p04) circle (1pt)
     (p10) circle (1pt)
     (p11) circle (1pt)
     (p12) circle (1pt)
     (p13) circle (1pt)
     (p14) circle (1pt)
     (p20) circle (1pt)
     (p21) circle (1pt)
     (p22) circle (1pt)
     (p23) circle (1pt)
     (p30) circle (1pt)
     (p31) circle (1pt)
     (p32) circle (1pt)
     (p40) circle (1pt)
     (p41) circle (1pt)
     (p50) circle (1pt);
\end{tikzpicture}
\caption{Construction~1.}
\label{fig-way1}
\end{minipage}%
\end{figure}

Cai and Yeung~\cite{Cai-Yeung-SNC-IT} have proved that there exists an $n$-dimensional SLNC with rate $\w$ and security level $r$ (here $n=\w+r$) on the network $G$ if and only if $\w+r\leq C_{\min}$, where $C_{\min}$ is the smallest minimum cut capacity between the source node and each sink node. We say a nonnegative integer pair $(\w, r)$ of rate $\w$ and security level $r$ is {\em achievable} if $\w+r\leq C_{\min}$, and the set of all the achievable pairs is called the {\em rate and security-level region}, depicted in~Fig.~\ref{fig-region}. In this section, we consider the ultimate problem of designing a family of local-encoding-preserving SLNCs for a flexible rate and a flexible security level, more precisely, a family of local-encoding-preserving SLNCs achieving all achievable pairs $(\w,r)$ in the rate and security level region.

To solve this problem, we combine Algorithm~1 in~\cite{part1} and Algorithms~\ref{algo-2}~and~\ref{algo-3} in the current paper, which respectively are designed to construct local-encoding-preserving SLNCs for a fixed security level and a flexible rate, for a fixed rate and a flexible security level, and for a fixed dimension and a flexible pair of rate and security level. By considering different combinations, we can provide multiple ways to construct a family of local-encoding-preserving SLNCs such that all pairs in the rate and security-level region are achieved. In the following, we present 3 possible constructions of such a family of local-encoding-preserving SLNCs.

\medskip

\textbf{Construction~1:} Start with a $C_{\min}$-dimensional linear network code $\mC_{C_{\min}}$. We first apply Algorithm~\ref{algo-3} to construct a $C_{\min}$-dimensional SLNC $\big(Q^{(C_{\min})}\big)^{-1}\cdot\,\mC_{\min}$ such that all the rate and security-level pairs on the line $\w+r=C_{\min}$ are achieved. Here, all the achievable pairs $(\w, r)$ with $\w+r=C_{\min}$ share the same SLNC $\big(Q^{(C_{\min})}\big)^{-1}\cdot\,\mC_{\min}$ and clearly the local-encoding-preserving property is guaranteed. By Theorem~\ref{thm_v-r-v-sl}, a field $\Fq$ of size
\begin{align*}
q>\max\big\{|T|,~|\mA_r|,~1\leq r \leq C_{\min}-1\big\}
\end{align*}
is sufficient for applying Algorithm~\ref{algo-3} to construct such an SLNC.

Next, for each achievable pair $(\w, r)$ on the line $\w+r=C_{\min}$, we start with the SLNC $\big(Q^{(C_{\min})}\big)^{-1}\cdot\,\mC_{\min}$ with rate $\w$ and security level $r$. Then, we apply Algorithm~1 in~\cite{part1} repeatedly to construct a family of local-encoding-preserving SLNCs with the fixed security level $r$ and rates decreasing one by one from~$\w$ to $0$. It follows from Theorem~6 in~\cite{part1} that a field $\Fq$ of size $q>\max\big\{|T|,~|\mA_r|\big\}$ is sufficient for applying Algorithm~1 in~\cite{part1} to construct such a family of local-encoding-preserving SLNCs. We remark that all the SLNCs in this family have the same local encoding kernels as $\big(Q^{(C_{\min})}\big)^{-1}\cdot\,\mC_{\min}$ at all the intermediate nodes. Therefore, the SLNCs in all the families with respect to distinct security levels $r$, $0\leq r \leq C_{\min}$ share a common local encoding kernel at each intermediate node, namely that all the SLNCs are local-encoding-preserving.

As such, we obtain a family of local-encoding-preserving SLNCs achieving all the pairs in the rate and security-level region. Furthermore, it follows from the above discussions that the field size
\begin{align}\label{field_size-Const-1}
\big|\Fq\big|>\max\big\{|T|,~|\mA_r|,~1\leq r \leq C_{\min}-1\big\}
\end{align}
is sufficient for constructing such a family of local-encoding-preserving SLNCs. By~\eqref{field_size-Const-1}, we see that with this method, there is no penalty on the field size (in terms of the best known lower bound~\cite{GY-SNC-Reduction}) for constructing such a family of SLNCs. Construction~1 is illustrated in Fig.~\ref{fig-way1}.

\medskip

\textbf{Construction~2:} Start with a $C_{\min}$-dimensional linear network code $\mC_{C_{\min}}$. We first apply Lemma~\ref{thm_decoding_condition} to construct local-encoding-preserving linear network codes $\mC_{n}$ for $0\leq n \leq C_{\min}$, and a base field $\Fq$ of size $q>|T|$ is sufficient (cf.~\cite{Yeung-book,Zhang-book}). Thus, we have obtained a family of local-encoding-preserving SLNCs with rate and security-level pairs $(n,0)$ for all $0\leq n \leq C_{\min}$. These pairs are all the nonnegative integer points on the line $r=0$.

Next, for each pair $(n,0)$, $1\leq n \leq C_{\min}$, we start with the SLNC $\mC_{n}$ (with the rate and security-level pair $(n,0)$). Then, we apply Algorithm~\ref{algo-2} repeatedly to construct a family of local-encoding-preserving SLNCs with the fixed rate~$n$ and security levels increasing one by one from $0$ to $C_{\min}-n$. Note that this construction is unnecessary for $n=0$ because the pair $(0,r)$ is always achievable for any $0\leq r \leq C_{\min}$. It follows from Theorem~9 that a field $\Fq$ of size
$q>\max\big\{|T|,~|\mA_{r}|,~1\leq r \leq C_{\min}-n \big\}$ is sufficient for applying Algorithm~\ref{algo-2} to construct such a family of local-encoding-preserving SLNCs. We note that all the SLNCs in this family have the same local encoding kernels as $\mC_n$ at all the intermediate nodes. Together with the assertion in Lemma~\ref{thm_decoding_condition} that all the linear network codes $\mC_{n}$, $0\leq n \leq C_{\min}$ share a common local encoding kernel at each intermediate node, we see that the SLNCs in all the families with respect to distinct rates $n$, $0\leq n \leq C_{\min}$ share a common local encoding kernel at each intermediate node, or equivalently, all the SLNCs are local-encoding-preserving.

As such, we have obtained a family of local-encoding-preserving SLNCs achieving all the pairs in the rate and security-level region, and by the above discussions, the field size
\begin{align*}
\big|\Fq\big|>\max\big\{|T|,~|\mA_r|,~1\leq r \leq C_{\min}-1\big\}
\end{align*}
is sufficient. Similar to Construction~1, there is no penalty on the field size (in terms of the best known lower bound~\cite{GY-SNC-Reduction}) for constructing such a family of SLNCs. Construction~2 is illustrated in Fig.~\ref{fig-way2}.

\begin{figure}[!t]
\centering
\begin{minipage}[b]{0.5\textwidth}
\centering
\begin{tikzpicture}[
>=stealth',
help lines/.style={dotted, thick},
axis/.style={<->},
important line/.style={thick, -latex},
important line 2/.style={thick},
connection/.style={thick, dashed}]
    \coordinate (y) at (0,6);
    \coordinate (x) at (6,0);
    \draw[<->] (y) node[above] {$r$} -- (0,0) node [below left]{$0$} --  (x) node[right]
    {$\w$};
    \path
    coordinate (p05) at (0,5)
    coordinate (p00) at (0,0)
    coordinate (p01) at (0,1)
    coordinate (p02) at (0,2)
    coordinate (p03) at (0,3)
    coordinate (p04) at (0,4)
    coordinate (p10) at (1,0)
    coordinate (p11) at (1,1)
    coordinate (p12) at (1,2)
    coordinate (p13) at (1,3)
    coordinate (p14) at (1,4)
    coordinate (p20) at (2,0)
    coordinate (p21) at (2,1)
    coordinate (p22) at (2,2)
    coordinate (p23) at (2,3)
    coordinate (p30) at (3,0)
    coordinate (p31) at (3,1)
    coordinate (p32) at (3,2)
    coordinate (p40) at (4,0)
    coordinate (p41) at (4,1)
    coordinate (p50) at (5,0);
    \draw[help lines] (p05) node [left] {$C_{\min}$} -- (p50) node[below]
    {$C_{\min}$};

    \draw[important line] (p50) node [left] {} -- (p40) node[below] {};
    \draw[important line] (p40) node [left] {} -- (p30) node[below] {};
    \draw[important line] (p30) node [left] {} -- (p20) node[below] {};
    \draw[important line] (p20) node [left] {} -- (p10) node[below] {};
    \draw[important line] (p10) node [left] {} -- (p00) node[below] {};

    \draw[important line 2] (p00) node [left] {} -- (p01) node[below] {};
    \draw[important line 2] (p01) node [left] {} -- (p02) node[below] {};
    \draw[important line 2] (p02) node [left] {} -- (p03) node[below] {};
    \draw[important line 2] (p03) node [left] {} -- (p04) node[below] {};
    \draw[important line 2] (p04) node [left] {} -- (p05) node[below] {};

    \draw[important line] (p10) node [left] {} -- (p11) node[below] {};
    \draw[important line] (p11) node [left] {} -- (p12) node[below] {};
    \draw[important line] (p12) node [left] {} -- (p13) node[below] {};
    \draw[important line] (p13) node [left] {} -- (p14) node[below] {};

    \draw[important line] (p20) node [left] {} -- (p21) node[below] {};
    \draw[important line] (p21) node [left] {} -- (p22) node[below] {};
    \draw[important line] (p22) node [left] {} -- (p23) node[below] {};

    \draw[important line] (p30) node [left] {} -- (p31) node[below] {};
    \draw[important line] (p31) node [left] {} -- (p32) node[below] {};

    \draw[important line] (p40) node [left] {} -- (p41) node[below] {};

    \draw[help lines] (p01) -- (p41);
    \draw[help lines] (p02) -- (p32);
    \draw[help lines] (p03) -- (p23);
    \draw[help lines] (p04) -- (p14);
     \filldraw [black]
     (p05) circle (1pt)
     (p00) circle (1pt)
     (p01) circle (1pt)
     (p02) circle (1pt)
     (p03) circle (1pt)
     (p04) circle (1pt)
     (p10) circle (1pt)
     (p11) circle (1pt)
     (p12) circle (1pt)
     (p13) circle (1pt)
     (p14) circle (1pt)
     (p20) circle (1pt)
     (p21) circle (1pt)
     (p22) circle (1pt)
     (p23) circle (1pt)
     (p30) circle (1pt)
     (p31) circle (1pt)
     (p32) circle (1pt)
     (p40) circle (1pt)
     (p41) circle (1pt)
     (p50) circle (1pt);
\end{tikzpicture}
\caption{Construction~2.}
\label{fig-way2}
\end{minipage}%
\centering
\begin{minipage}[b]{0.5\textwidth}
\centering
\begin{tikzpicture}[
>=stealth',
help lines/.style={dotted, thick},
axis/.style={<->},
important line/.style={thick, latex-},
important line 2/.style={thick},
connection/.style={thick, dashed}]
    \coordinate (y) at (0,6);
    \coordinate (x) at (6,0);
    \draw[<->] (y) node[above] {$r$} -- (0,0) node [below left]{$0$} --  (x) node[right]
    {$\w$};
    \path
    coordinate (p05) at (0,5)
    coordinate (p00) at (0,0)
    coordinate (p01) at (0,1)
    coordinate (p02) at (0,2)
    coordinate (p03) at (0,3)
    coordinate (p04) at (0,4)
    coordinate (p10) at (1,0)
    coordinate (p11) at (1,1)
    coordinate (p12) at (1,2)
    coordinate (p13) at (1,3)
    coordinate (p14) at (1,4)
    coordinate (p20) at (2,0)
    coordinate (p21) at (2,1)
    coordinate (p22) at (2,2)
    coordinate (p23) at (2,3)
    coordinate (p30) at (3,0)
    coordinate (p31) at (3,1)
    coordinate (p32) at (3,2)
    coordinate (p40) at (4,0)
    coordinate (p41) at (4,1)
    coordinate (p50) at (5,0);

    \draw[important line] (p40) node [left] {} -- (p50) node[below] {};
    \draw[important line] (p30) node [left] {} -- (p40) node[below] {};
    \draw[important line] (p20) node [left] {} -- (p30) node[below] {};
    \draw[important line] (p10) node [left] {} -- (p20) node[below] {};
    \draw[important line] (p00) node [left] {} -- (p10) node[below] {};

    \draw[important line 2] (p05) node [left] {$C_{\min}$} --  (p50) node[below] {$C_{\min}$};
    \draw[important line 2] (p04) -- (p40);
    \draw[important line 2] (p03) -- (p30);
    \draw[important line 2] (p02) -- (p20);
    \draw[important line 2] (p01) -- (p10);

    \draw[help lines] (p10) -- (p14);
    \draw[help lines] (p20) -- (p23);
    \draw[help lines] (p30) -- (p32);
    \draw[help lines] (p40) -- (p41);
    \draw[help lines] (p01) -- (p41);
    \draw[help lines] (p02) -- (p32);
    \draw[help lines] (p03) -- (p23);
    \draw[help lines] (p04) -- (p14);
     \filldraw [black]
     (p05) circle (1pt)
     (p00) circle (1pt)
     (p01) circle (1pt)
     (p02) circle (1pt)
     (p03) circle (1pt)
     (p04) circle (1pt)
     (p10) circle (1pt)
     (p11) circle (1pt)
     (p12) circle (1pt)
     (p13) circle (1pt)
     (p14) circle (1pt)
     (p20) circle (1pt)
     (p21) circle (1pt)
     (p22) circle (1pt)
     (p23) circle (1pt)
     (p30) circle (1pt)
     (p31) circle (1pt)
     (p32) circle (1pt)
     (p40) circle (1pt)
     (p41) circle (1pt)
     (p50) circle (1pt);
\end{tikzpicture}
\caption{Construction~3.}
\label{fig-way3}
\end{minipage}
\end{figure}

\medskip

\textbf{Construction~3:} Start with a $C_{\min}$-dimensional linear network code $\mC_{C_{\min}}$. Similar to Construction~2, we first apply Lemma~\ref{thm_decoding_condition} to construct a family of local-encoding-preserving SLNCs with rate and security-level pairs $(n,0)$, $0\leq n \leq C_{\min}$, where the field size $|\Fq|>|T|$ is sufficient.

Next, for each pair $(n,0)$, $1\leq n \leq C_{\min}$, we start with the SLNC $\mC_{n}$ and apply Algorithm~\ref{algo-3} to construct an $n$-dimensional SLNC $\big(Q^{(n)}\big)^{-1}\cdot\,\mC_{n}$ achieving all the nonnegative rate and security-level pairs $(\w,r)$ with $\w+r=n$. It follows from Theorem~\ref{thm_v-r-v-sl} that a base field $\Fq$ of size $q>\max\big\{|T|,~|\mA_r|,~1\leq r \leq n-1\big\}$ is sufficient for applying Algorithm~\ref{algo-3} to construct such an SLNC. Furthermore, by Proposition~\ref{prop}, the SLNC $\big(Q^{(n)}\big)^{-1}\cdot\,\mC_{n}$ has the same local encoding kernels as $\mC_{n}$ at all the intermediate nodes. By Lemma~\ref{thm_decoding_condition}, all $\mC_{n}$, $0\leq n \leq C_{\min}$ share a common local encoding kernel at each intermediate node. We thus see that all the SLNCs $\big(Q^{(n)}\big)^{-1}\cdot\,\mC_{n}$, $1\leq n \leq C_{\min}$ share a common local encoding kernel at each intermediate node, namely that all the SLNCs are local-encoding-preserving.

As such, we have obtained a family of local-encoding-preserving SLNCs achieving all the pairs in the rate and security-level region, and by the above discussions, the field size
\begin{align*}
\big|\Fq\big|>\max\big\{|T|,~|\mA_r|,~1\leq r \leq C_{\min}-1\big\}
\end{align*}
is sufficient. Again, we see that no penalty on the field size (in terms of the best known lower bound~\cite{GY-SNC-Reduction}) exists for constructing such a family of SLNCs. Construction~3 is illustrated in Fig.~\ref{fig-way3}.

\section{Conclusion}\label{Sec_conclusion}

In this Part~II of a two-part paper, we continue the studies of local-encoding-preserving secure network coding in Part~\Rmnum{1}~\cite{part1}. We first tackle the problem of local-encoding-preserving secure network coding for a fixed rate and a flexible security level. We develop a novel approach for designing a family of local-encoding-preserving SLNCs with a fixed rate and the security level ranging from $0$ to the maximum possible. Our approach, which increases the dimension of the code (equal to the sum of the rate and the security level) step by step, is totally different from all the previous approaches for related problems which decreases the dimension of the code. A polynomial-time algorithm is presented for efficient implementation of our approach. We also prove that our approach does not incur any penalty on the required field size for the existence of SLNCs in terms of the best known lower bound by Guang and Yeung~\cite{GY-SNC-Reduction}, and has a constant storage cost that is independent of the dimension of the SLNC.

We further tackle the problem of local-encoding-preserving secure network coding for a fixed dimension. We develop another novel approach for designing an SLNC that can be applied for all the rate and security-level pairs with the fixed dimension. Clearly, the local-encoding-preserving property is guaranteed since only one SLNC is applied for all such pairs. Based on this approach, we present a polynomial-time algorithm for implementation and prove that there is no penalty on the required field size for the existence of SLNCs (also in terms of the best known lower bound by Guang and Yeung~\cite{GY-SNC-Reduction}).

The code constructions presented in Part~I and the current paper can be applied individually. At the end of the paper, we
show that by combining these constructions in suitable ways, they can be used for solving
the ultimate problem of local-encoding-preserving secure network coding for the whole rate and security-level region.


\numberwithin{thm}{section}
\appendices

\end{document}